\documentclass[iop,apj,numberedappendix]{emulateapj}

\pdfoutput=1

\usepackage{graphicx}

\usepackage{algorithm2e}
\SetAlgoSkip{medskip}
\SetAlgoInsideSkip{medskip}
\DontPrintSemicolon

\usepackage[colorlinks,urlcolor=blue,citecolor=blue,linkcolor=blue,backref,breaklinks]{hyperref}
\usepackage[all]{hypcap}

\graphicspath{{figures/}}

\newcommand{\Ss}{\S\S}

\newcommand{\sci}[2]{\mbox{$#1 \times 10^{#2}$}}

\newcommand{\rmscriptsize}[1]{\textrm{\scriptsize{#1}}}
\newcommand{\SUB}[2]{\ensuremath{#1_{\rmscriptsize{#2}}}}
\newcommand{\SUBB}[3]{\ensuremath{#1_{#2,\,\rmscriptsize{#3}}}}

\newcommand{\GJ}[1]{\SUB{#1}{\textsc{gj}}}
\newcommand{\PC}[1]{\SUB{#1}{pc}}    

\newcommand{\RNS}{\SUB{R}{\textsc{ns}}}

\newcommand{\jm}{\SUB{j}{m}}

\newcommand{\lambdaC}{\SUB{\lambdabar{}}{\textsc{c}}}
\newcommand{\alphaF}{\ensuremath{\alpha_f}}    

\newcommand{\chiA}{\ensuremath{\chi_{a}}}    
\newcommand{\psiA}{\ensuremath{\psi_{a}}}    

\newcommand{\rhoC}{\SUB{\rho}{c}}

\newcommand{\epsilonG}{\ensuremath{\epsilon_{\gamma}}}    
\newcommand{\epsilonP}{\ensuremath{\epsilon_{\pm}}}    

\newcommand{\ESC}[1]{\SUB{#1}{esc}}
\newcommand{\epsilonGesc}{\ensuremath{\epsilon_{\gamma,\,\rmscriptsize{esc}}}}

\newcommand{\SYN}[1]{\SUB{#1}{syn}}
\newcommand{\CR}[1]{\SUB{#1}{\textsc{cr}}}

\newcommand{\CRsyn}[1]{\SUB{#1}{\textsc{cr}-syn}}

\newcommand{\epsilonGSYN}{\SUBB{\epsilon}{\gamma}{syn}}
\newcommand{\epsilonGCR}{\SUBB{\epsilon}{\gamma}{\textsc{cr}}}
\newcommand{\epsilonGRICS}{\SUBB{\epsilon}{\gamma}{\textsc{RICS}}}

\newcommand{\GAP}[1]{\SUB{#1}{gap}}
\newcommand{\GAParg}[2]{\ensuremath{#1_{#2,\, \rmscriptsize{gap}}}}

\newcommand{\lP}{\ensuremath{l_\pm}}
\newcommand{\lPe}{\ensuremath{l_\pm^{\rmscriptsize{e}}}}
\newcommand{\lPeGAP}{\ensuremath{l_{\pm,\, \rmscriptsize{gap}}^{\rmscriptsize{e}}}}

\shorttitle{Pair multiplicity in young pulsars}
\shortauthors{Timokhin \& Harding}

\begin{document}

\title{On the polar cap cascade pair multiplicity of young pulsars}

\author{A.~N.~Timokhin$^{1,2}$ and A.~K.~Harding$^{1}$}

\affil{%
  $^1$Astrophysics Science Division, NASA/Goddard Space Flight Center,
  Greenbelt, MD 20771, USA\\
  $^2$University of Maryland, College Park (UMDCP/CRESST), College Park, MD 20742, USA}

\email{andrey.timokhin@nasa.gov}

\date{Received ; accepted ; published }

\begin{abstract}
  We study the efficiency of pair production in polar caps of young
  pulsars under a variety of conditions to estimate the maximum
  possible multiplicity of pair plasma in pulsar magnetospheres. We
  develop a semi-analytic model for calculation of cascade
  multiplicity which allows efficient exploration of the parameter
  space and corroborate it with direct numerical simulations.  Pair
  creation processes are considered separately from particle
  acceleration in order to assess different factors affecting cascade
  efficiency, with acceleration of primary particles described by
  recent self-consistent non-stationary model of pair cascades.  We
  argue that the most efficient cascades operate in the curvature
  radiation/synchrotron regime, the maximum multiplicity of pair
  plasma in pulsar magnetospheres is $\sim\mbox{few}\times10^5$.  The
  multiplicity of pair plasma in magnetospheres of young energetic
  pulsars weakly depends on the strength of the magnetic field and the
  radius of curvature of magnetic field lines and has a stronger
  dependence on pulsar inclination angle. This result questions
  assumptions about very high pair plasma multiplicity in theories of
  pulsar wind nebulae.
\end{abstract}

\keywords{  
  acceleration of particles --- 
  plasmas --- 
  pulsars: general --- 
  stars: neutron}

\section{Introduction}

The idea that production of electron-positron pairs in magnetospheres
of rotation-powered pulsars is intimately connected with their
activity had been proposed by \citet{Sturrock71} only a few years
after the discovery of pulsars \citep{HewishBell1968}. Since then it
has become an integral part of the standard pulsar model and today,
there is little doubt that an active rotationally powered pulsar
produces electron-positron plasma.  Although the pulsar emission
mechanism(s) is still not yet identified, there is strong empirical
evidence that pulsars stop emitting radio waves when pair formation
ceases -- the threshold for pair formation roughly corresponds to the
``death line'' in pulsar parameter space, what was already noted by
\citet{Sturrock71}.  Furthermore, the narrow peaks in many pulsar
high-energy light curves \citep{Abdo2010_FirstPSRCatalog} require
pervasive screening of the whole magnetosphere by pair plasma, except
in narrow accelerator gaps \citep[e.g.][]{Watters2009,
  Pierbattista2015}.  Understanding pair plasma generation in pulsar
magnetospheres is therefore of crucial importance for developing
pulsar emission models.

In the standard pulsar model, initially proposed by \citet{GJ} and
\citet{Sturrock71}, the magnetosphere is filled with dense pair plasma
which screens the accelerating electric field everywhere except some
small zones which are responsible for particle acceleration and
emission.  Pair plasma is primarily created via conversion of $\gamma$
rays in the strong magnetic field near the polar caps (PCs).  Pair
production in the PCs is a ``cornerstone'' of the standard model --
without dense plasma produced at the PCs, at the base of open magnetic
field lines, the magnetosphere would have large volumes with
unscreened electric field, as pair creation in e.g. outer gaps
\citep{Cheng/Ruderman76} cannot screen the electric field over the
rest of the magnetosphere.

Charge starvation \citep{AronsScharlemann1979} or vacuum gaps
\citep{RudermanSutherland1975} at the polar cap, when the number
density of charged particles is not enough to screen the electric
field, leads to formation of accelerating zone(s).  Some charged
particles enter this zone, are accelerated to very high energies and
emit gamma-rays which are absorbed in the ultrastrong magnetic field,
creating electron-positron pairs. The pairs, being relativistic, can
also emit pair producing photons and so the avalanche develops until
photons emitted by the last generation of pairs can no longer produce
pairs and escape the magnetosphere.

The pair plasma created by pulsars flows out of the magnetosphere
along open magnetic field lines and provides the radiating particles
for the surrounding Pulsar Wind Nebulae (PWNe).  Models of PWNe depend
(at least) on the density of the plasma, what produces the observed
synchrotron and inverse Compton emission.  Estimates of the pair
multiplicity (the number of pairs produced by each primary accelerated
particle) needed to account for the emission from the Crab pulsar wind
nebula (PWN) range from about $10^5 - 10^6$ \citep{deJager1996} up to
$10^7$ \citep[e.g.][]{BucciantiniAronsAmato2011}; for the Vela PWN the
multiplicity is estimated to be about $10^5$ \citep{deJager2007}.
PWNe therefore give the most compelling evidence for pair production
and pair cascades in at least young energetic pulsar magnetospheres.

Although PWNe are observed only around young pulsars ($<$ a few times
$10^4$ yrs), evidence for pairs, at least for high plasma densities
larger than those provided by primary particles, can also be found in
older pulsars.  Synchrotron absorption models for the eclipse in the
double pulsar system PSR J0737-3039 \citep{Arons2005, Lyutikov2004}
require a pair multiplicity of around $10^6$ for the recycled 22 ms
pulsar in that system.

The cascade process in pulsar polar caps has been the subject of
extensive numerical as well as analytical studies
\citep[e.g.][]{Daugherty/Harding82,GurevichIstomin1985,Zhang2000,
  Hibschman/Arons:pair_multipl::2001,Hibschman/Arons:pair_production::2001,Medin2010}.
The pair plasma multiplicity obtained in these studies was
significantly lower than estimates of pair plasma multiplicity in
PWNe, as it did not exceed $\sim$few$\times10^4$.  Most of those works
considered pair creation together with the particle acceleration which
makes these analyses dependent on the acceleration model considered.
These studies also assumed steady, time-independent acceleration of
the primary particles.  However, recent studies by
\citet{Timokhin2010::TDC_MNRAS_I,TimokhinArons2013} have found that
pulsar polar cap pair cascades are not time-steady in the general case
of arbitrary current, particularly those required by global
magnetosphere models
\citep[e.g][]{CKF,Timokhin2006:MNRAS1,Spitkovsky:incl:06,Kalapotharakos2009}.

In this paper we study the question of what is the maximum pair multiplicity
achievable in pulsar polar cap cascades and under which circumstances is it
achieved.  In contrast to previous pair cascade studies, we take a multistep
approach.  We consider the physical processes in pair cascades and particle
acceleration models separately in order to clearly set apart different factors
influencing the efficiency of pair cascades.  We first assess how each of the
microscopic processes affects the final multiplicity and the pulsar parameter
ranges that result in the largest possible pair multiplicity.  Then, we employ
the most recent model of non steady-state particle acceleration in pulsar polar
caps and derive a simple analytical estimate for the maximum energy of particles
accelerated in a non-stationary cascade. One on the most important results of
our study is a strong upper limit on pair plasma multiplicity in pulsars.

We limit ourselves to the case of cascades at the polar caps of young%
\footnote{Pulsars with strong polar cap cascades should have potential drop over
  the polar cap well in excess of the pair formation threshold as well as large
  magnetic fields $B\gtrsim10^{11}$~G and short rotational periods, so that
  particle acceleration happens over a short distance and cascade develops in
  the region with strong magnetic field. The best single parameter selecting
  pulsars with such properties is the small characteristic age
  $\tau=P/2\dot{P}$. A detailed discussion of the pulsar parameter range where
  approximations used in this paper are formally applicable is given at the end
  of \S~\ref{sec:energy-primary}.}
pulsars as from previous theoretical studies of polar cap cascades, such pulsars
are expected to be the most efficient pair producers.  We rely on results of
previous cascade studies in our choice of the specific cascade process, namely
cascades initiated by curvature radiation of primary particles.

The plan of the paper is as follows.  In
\S~\ref{sec:cascades-overview} we briefly discuss the efficiency of
cascades in general and give an overview of the most efficient cascade
process in pulsar polar caps.  In subsequent sections we consider in
detail all physical processes in such cascades.
\Ss\ref{sec:photon-absorption}--\ref{sec:mult-semi-analytical}, the
largest part of this paper, are devoted to development of a simple
semi-analytical model for estimation of pair production efficiency in
polar caps of young pulsars. This model allows efficient exploration
of pulsar parameter space and helps to clarify the main factors
affecting cascade multiplicity.  Then, in \S\ref{sec:mult-numerical}
we use direct numerical simulations of PC cascades in a Crab-like
pulsar to show that predictions of our semi-analytical model are
indeed correct. We discuss uncertainties of current pulsar models in
\S~\ref{sec:flux-primary} and summarize and discuss our findings and
their implications in the Discussion.

\section{Physics of polar cap cascades: an overview }
\label{sec:cascades-overview}

An electron-positron cascade can be thought of as a process of
splitting the energy of primary particles into the energies of
secondary particles.  The maximum multiplicity, the number of
secondary particles for each primary particle, of an ideal cascade
initiated by a single primary particle with energy $\epsilon_p$ would
be
\begin{equation}
  \label{eq:kappa_max}
  \SUB{\kappa}{max}\simeq{}2\, \frac{\epsilon_p}{\epsilonGesc}\,.
\end{equation}
where $\epsilonGesc$ is the maximum energy of photons escaping from
the cascade (or the minimum energy of pair producing photons).  Not
all of the primary particle energy goes into pair producing photons,
and pairs created in the cascade do not radiate all of their energy
into the next generation of pair producing photons.  Hence,
$\SUB{\kappa}{max}$ is only the theoretical upper limit on the
multiplicity of a real cascade.  The total pair yield of the cascade
is a combination of four factors: (a) number of primary particles, (b)
initial energy of primary particle -- the higher the energy the more
pairs can be produced, (c) threshold for pair formation -- the lower
the threshold the higher the multiplicity, (d) efficiency of splitting
the energy of primary particles into pairs -- the higher the fraction
of particle energy going into pair creating photons, as opposed to the
final kinetic energy of the particles and photons below the pair
formation threshold, the higher the multiplicity.

In young, fast rotating pulsars, the electric field in the polar cap
acceleration zones is strong and primary particles can be accelerated
up to very high Lorentz factors, $\gamma\gtrsim{}10^7$.  At these
energies the most effective radiation process is curvature radiation
(CR).  CR efficiency grows rapidly with the particle energy and for
young pulsars becomes the dominant emission mechanism for primary
particles.  For secondary particles, which are substantially less
energetic than the primary ones, the primary way to create pair
producing photons is via synchrotron radiation.  In
\S~\ref{sec:reson-inverse-compt} we argue that although another
possible emission mechanism for secondary particles -- Resonant
Inverse Compton Scattering (RICS) of soft X-ray photons emitted by the
NS -- can generate pair producing photons in pulsars with magnetic
field $\gtrsim\sci{3}{11}$~G, that channel never becomes the dominant
one for pair production in young pulsars, at best resulting in pair
multiplicity comparable to the one of the synchrotron channel.  Hence,
in young pulsars which are expected to have the highest multiplicity
pair plasmas, the polar cap cascades should operate primarily in the
CR-synchrotron regime -- all known studies of cascades in pulsar polar
cap agree on this point.  In this paper we study in detail
CR-synchrotron cascades.  The resulting multiplicities will be good
estimates for a wide range of pulsar parameters, however, as we
consider only synchrotron radiation of secondary particles, for
pulsars with magnetic field $\gtrsim\sci{3}{12}$~G our analysis might
underestimate the cascade multiplicity by a factor of $\sim2$, see
\S~\ref{sec:reson-inverse-compt}.

\begin{figure}
\label{fig:full_cascade}
\includegraphics[clip,width=\columnwidth]{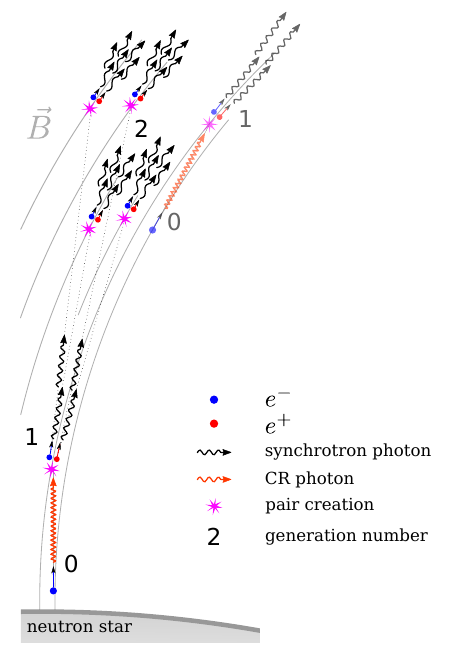}
\caption{Schematic representation of electron-positron cascade in the
  polar cap of a young pulsar, see text for description.}
\end{figure}

Fig.~\ref{fig:full_cascade} gives a schematic overview of
electron-positron cascade development in polar cap regions of young
pulsars.  Shown are the first two generations in a cascade initiated
by a primary electron.  Primary electrons emit CR photons (almost)
tangent to the magnetic field lines; primary electrons and CR photons
are generation 0 particles in our notation.  Magnetic field lines are
curved and the angle between the photon momentum and the magnetic
field grows as the photon propagates further from the emission point.
When this angle becomes large enough, photons are absorbed and each
photon creates an electron-positron pair -- generation 1 electron and
positron.  The pair momentum is directed along the momentum of the
parent photon and at the moment of creation, the particles have
non-zero momentum perpendicular to the magnetic field.  They radiate
this perpendicular momentum almost instantaneously via synchrotron
radiation and then move along magnetic field lines.  Although these
secondary particles are relativistic, their energy is much lower than
that of the primary electron and their curvature photons cannot create
pairs.  After the emission of synchrotron photons, secondary particles
(generation 1 and higher) no longer contribute to cascade development.
Generation 1 photons (synchrotron photons produced by the generation 1
particles) are also emitted (almost) tangent to the magnetic field
line -- as the secondary particles are relativistic -- and propagate
some distance before acquiring the necessary angle to the magnetic
field and creating generation 2 pairs.  These pairs in their turn
radiate their perpendicular momentum via synchrotron radiation,
emitting generation 2 photons.  The cascade initiated by a single CR
photon stops at a generation where the energy of synchrotron photons
falls below $\epsilonGesc$.

Only primary particles emit pair producing photons as they move along
the field lines; all secondary particles emit pair producing photons
just after their creation.  The cascade development can be thus
divided into two parts: (i) primary particles emit CR photons as they
move along magnetic field lines and (ii) each CR photon gives rise to
a synchrotron cascade, when synchrotron photons create a successive
generation of pairs which emit the next generation synchrotron photons
at the moment of creation.  This division goes between generation 0
and all subsequent cascade generations.

In the following sections we analyze all four factors regulating the
yield of CR-synchrotron cascades listed at the beginning of this
section (in reverse order, from d to a).  In
\Ss\ref{sec:photon-absorption}--\ref{sec:mult-semi-analytical} we
develop a simple semi-analytical models for calculation of
multiplicity of strong polar caps cascades initiated by a primary
electron which we then compare with detailed numerical computation
described in \S\ref{sec:mult-numerical}.  We start with gamma-ray
absorption in a strong magnetic field in
\S\ref{sec:photon-absorption}; then we discuss the efficiency of the
synchrotron cascade in \S\ref{sec:synchrotron-cascade}.  Curvature
radiation is considered in \S\ref{sec:curvature-radiation} and in
\S\ref{sec:mult-CR-synch} we discuss the multiplicity of a
CR-synchrotron cascade initiated by a single particle of given
energy. This covers items c and d from the list of factors affecting
cascades efficiency.  In \S\ref{sec:particle-acceleration} we address
item b from the list - energy of primary particles initiating the
cascade. \Ss\ref{sec:mult-semi-analytical} and
\ref{sec:mult-numerical} are devoted to the total multiplicity of a
CR-synchrotron cascade when particle acceleration is taken into
account - in \S\ref{sec:mult-semi-analytical} we present results for
cascade multiplicities from a semi-analytical model and in
\S\ref{sec:mult-numerical} we present results of detailed numerical
simulations of cascades in a Crab-like pulsar.  In
\S\ref{sec:reson-inverse-compt} we discuss the role of RICS in polar
caps cascades and argue that considering only CR-synchrotron cascades
gives us an adequate estimate of the pair multiplicity in young
pulsars.  Finally, in \S\ref{sec:flux-primary}, we address item a --
the mean flux of primary particles and the total yield of a
CR-synchrotron cascade in an energetic pulsar. We argue that this is
the most important factor regulating pair yield in energetic
pulsars. Despite the uncertainty in determining the mean flux of
primary particles, we can set a rather strict upper limit on pair
multiplicity in pulsars.

\section{Photon absorption in the magnetic field}
\label{sec:photon-absorption}

\subsection{Opacity for $\gamma-B$ pair production}
\label{sec:opacity-gamma-b}

\begin{figure*}[t]
\includegraphics[clip,width=\textwidth]{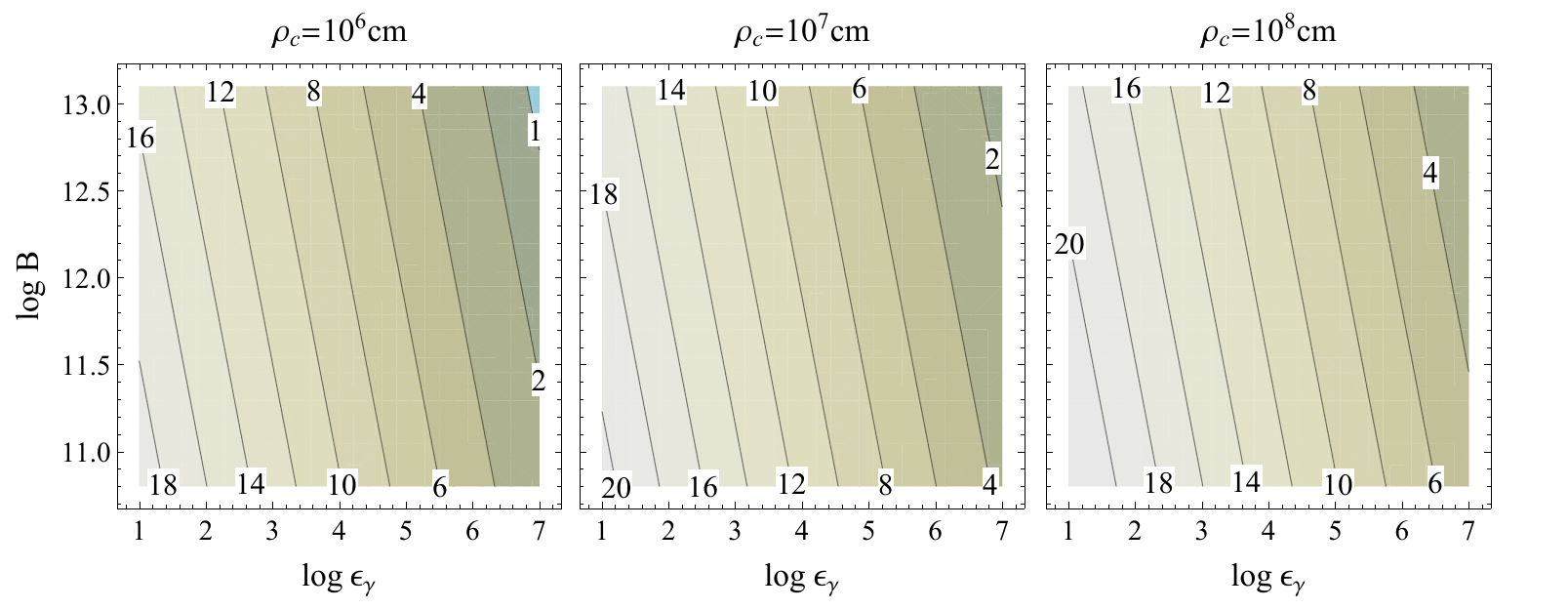}
\caption{Contour plot of $1/\chiA$ as a function of the logarithms of
  the magnetic field strength $B$ in Gauss, and photon energy
  $\epsilonG$ normalized to the electron rest energy, for three values
  of the radius of curvature of magnetic field lines
  $\rhoC=10^6, 10^7, 10^8\mbox{cm}$. $1\chiA$ values shown on this
  plot are calculated from eq.~(\ref{eq:chiA}) and are not corrected
  for the kinematic threshold (see text).}
\label{fig:inverse_chi}
\end{figure*}

\begin{figure*}[t]
\includegraphics[clip,width=\textwidth]{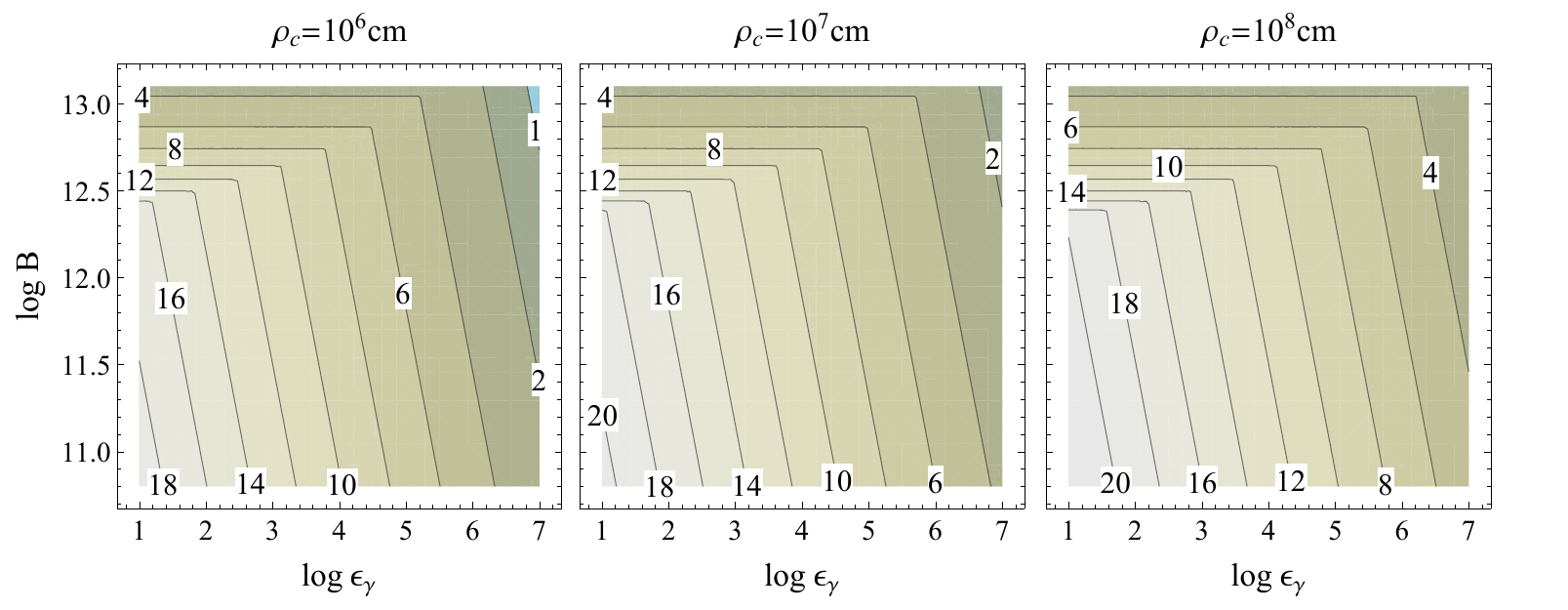}
\caption{Contour plot of $1/\chiA$ corrected for kinematic threshold
  according to eq.~(\ref{eq:chiA_limited}). These values of $1/\chiA$
  are used in all calculations within semi-analytic model.  Notations
  are the same as in Fig.~\ref{fig:inverse_chi}.}
\label{fig:inverse_chi_corrected_rho6}
\end{figure*}

The opacity for single photon pair production in strong magnetic field
is \citep{Erber1966}
\begin{equation}
  \label{eq:alphaB}
  \alpha_B(\epsilonG,\psi) = 0.23\,\frac{\alphaF}{\lambdaC}\:b\,\sin\psi\,\exp\left(-\frac{4}{3\chi}\right) 
\end{equation}
where $b\equiv{}B/B_q$ is the local magnetic field strength $B$
normalized to the critical quantum magnetic field 
$B_q=e/\alphaF\lambdaC^2=\sci{4.41}{13}$~G, 
$\psi$ is the angle between
the photon momentum and the local magnetic field, 
$\alphaF=e^2/\hbar{}c\approx{}1/137$
is the fine structure constant, and
$\lambdaC = \hbar/m c = \sci{3.86}{-11}$~cm
is the reduced Compton wavelength. The parameter $\chi$ is defined as
\begin{equation}
\label{eq:chi_def}
  \chi \equiv \frac{1}{2}\, \epsilonG b\, \sin\psi\,,
\end{equation}
where $\epsilonG$ is the photon energy in units of $m_ec^2$. For
convenience from here on, all particle and photon energies will be
quoted in terms of $m_ec^2$.  The optical depths for pair creation by
a high energy photon in a strong magnetic field after propagating distance
$l$ is
\begin{equation}
\label{eq:tau_general}
\tau(\epsilonG,l)=\int_0^l\alpha_B(\epsilonG,\psi(x))\,dx\,, 
\end{equation}
where integration is along the photon's trajectory.

Expression~(\ref{eq:alphaB}) is accurate if the magnetic field is
small compared to the critical field $B_q$, $b<0.2$ suffices, and if
$\epsilonG\sin\psi>2$ so that the created pair is in a high Landau
level; pair production into low Landau levels and for higher magnetic
fields has been discussed in \citet{DaughertyHarding1983} and
\citet{HardingBaringGonthier1997}.  For most pulsars, the magnetic
field in polar cap regions is smaller that $B_q/3$. In this paper we
study cascades in pulsars with ``normal'' magnetic fields and so we
neglect high-field effects.

Expression~(\ref{eq:alphaB}) becomes inaccurate when pairs are created
at low Landau levels, near the pair formation threshold, it
overestimates the opacity and even formally allows pair formation
below the kinematic threshold $\epsilonG\sin\psi=2$.  However, pairs
created at low Landau levels will not give rise to strong cascades, as
their perpendicular energy will be too low to emit pair-producing
synchrotron photons.  Hence, for the case of strong cascades, accurate
treatment of pair formation near the kinematic threshold is not
necessary.  Throughout this paper we use Erber's
approximation~(\ref{eq:alphaB}) in all our analytical calculations but
introduce a cut off at the kinematic threshold $\epsilonG\sin\psi=2$
as described at the end of this section, thus taking into account the
cessation of pair formation below the the kinematic threshold.

If the mean free path (mfp) of a pair-producing photon $l_\gamma$ is
comparable to or larger than the characteristic scale of the magnetic
field variation $L_B$, this photon will not initiate a strong cascade
with the same emission processes by which it was produced. The reason
for this is as follows.  The opacity for $\gamma{}B$ pair production
exponentially depends on the magnetic field strength and photon energy
via $\chi$.  The energy of the next generation photon will be smaller
than that of the primary one, and, because the primary photon has
already traveled the distance over which the magnetic field has
substantially declined, the magnetic field along the next generation
photon's trajectory will be substantially weaker than that along the
primary photon's trajectory%
\footnote{In the polar cap, photons are emitted by ultrarelativistic
  particles moving along magnetic field lines. At emission points
  photons are almost tangent to field lines and the differences in
  initial photon pitch angles for different generations can be
  neglected.}.
The next generation photon's mfp will be much larger than than that of
the primary photon, and, even if this secondary photon will be
absorbed, it will be the last cascade generation.  Hence, in a strong
cascade, for all but the last generation photons, $l_\gamma\ll{}L_B$.
A reasonable estimate for $L_B$ would be the distance of the order of
the NS radius $\RNS$ as any global NS magnetic field decays with the
distance as $(r/\RNS)^{-\delta}$, $\delta\ge3$.  Very localized, sun
spot like magnetic fields, are in our opinion of no importance for the
general pulsar case as the probability of such a ``spot'' to lie at
the polar cap should be rather low, i.e. most pulsars should be able
to produce plasma in a more or less regular magnetic field. A dipole
field, $\delta=3$, is often considered as a reasonable assumption for
a general pulsar model. Pure dipole field, however, seems to be a too
idealized approximation, as even if the NS field is a pure dipole, it
will be slightly disturbed by the currents flowing in the
magnetosphere.  In general, near dipole magnetic fields with different
curvatures of magnetic field lines should be examined in cascade
models.

We consider strong cascades with large multiplicities, where, as
argued above, photons propagate distances much shorter than the
characteristic scale of the magnetic field variation, so we assume
that in the region where most of the pairs are produced the magnetic
field is constant.  The radius of curvature of magnetic field lines
$\rhoC$ is not smaller than $L_B$, and as $l_\gamma\ll{}L_B$ the angle
$\psi$ is always small, the approximation $\sin\psi\approx\psi$ is
very accurate.  For photons emitted tangent to the magnetic field
line, $dx=\rhoC d\psi$.  In our approximation both $b$ and $\rhoC$ are
constants. From eq.~(\ref{eq:chi_def}) we have
$\psi=2\chi/\epsilonG{}b$, and substituting it into
eq.~(\ref{eq:tau_general}) we can express the optical depth $\tau$ to
pair production as an integral over $\chi$
\begin{equation}
  \label{eq:tau_OTS_general}
  \tau(\epsilonG,l) = A_\tau\, \frac{\rhoC}{\epsilonG^2 b}\;
  \int_0^{\chi(\epsilonG, \psi(l))} \chi
  \exp\left(-\frac{4}{3\chi}\right)\, d\chi\,,
\end{equation}
where
$A_\tau\equiv{}0.92 \alphaF/\lambdaC\approx\sci{1.74}{8}\;\mbox{cm}^{-1}$.

Integrating eq.~(\ref{eq:tau_OTS_general}) over $\chi$ by parts two
times we can get an expression for $\tau$ in terms of elementary
functions and the exponential integral function $\mbox{Ei}$:
\begin{eqnarray}
  \label{eq:tau_OTS}
  \tau(\chi) & = & A_\tau\, \frac{\rhoC}{\epsilonG^2 b}\times\nonumber\\
  & & \left[
      \frac{\chi^2}{2}\left(1-\frac{4}{3\chi}\right) e^{-\frac{4}{3\chi}} - 
      \frac{8}{9}\,\mbox{Ei}\left(-\frac{4}{3 \chi }\right)
      \right]\,.
\end{eqnarray}
$\mbox{Ei}(z)$ is a widely used special function, defined as
$\mbox{Ei}(z)=-\int_{-z}^\infty\exp(-t)/t\,dt$.  There are efficient
numerical algorithms for its calculation implemented in many numerical
libraries and scientific software tools; using eq.~(\ref{eq:tau_OTS})
for calculation of the optical depths will result in much more
efficient numerical codes than direct integration of
eq.~(\ref{eq:tau_OTS_general}).

As the optical depth to pair formation grows exponentially with $\chi$
(and distance), for analytical estimates it is reasonable to assume
that all photons are absorbed when they have traveled the distance
$l_\gamma$ such that $\tau(l_\gamma)=1$.  We denote the value of
$\chi$ when the optical depth reaches 1 as $\chiA:$
\begin{equation}
  \label{eq:chiA}
  \chiA: \tau(\chiA)=1\,.
\end{equation}
In all our computations we will use $\chiA$ as the value of the
parameter $\chi$ at the point of the photon's absorption.  $\chiA$ is
a solution of the non-linear equation~(\ref{eq:chiA}) with $\tau$
given by eq.~(\ref{eq:tau_OTS}).  Because of the exponential
dependence of $\tau$ on $1/\chi$ it is to be expected that $1/\chiA$
should have a close to linear dependence on logarithms of $\epsilonG$,
$b$, and $\rhoC$.  We solved equation~(\ref{eq:chiA}) numerically for
different values of $\epsilonG$, $b$, and $\rhoC$ and, indeed, the
inverse quantity $1/\chiA$ depends almost linearly on $\log\epsilonG$,
$\log{}b$, and $\log\rhoC$ in a wide range of these parameters.
Making a modest size table of $1/\chiA$ values one can later use
piecewise interpolation to find a particular value of $\chiA$ quite
accurately.

In Fig.~\ref{fig:inverse_chi} we plot contours of $1/\chiA$ as
functions of $\log(\epsilonG)$ and $\log(B)$ for three different
values for the radius of curvature of the magnetic field lines. We
want to point out that $1/\chiA$ differs from the value $1/\chiA=15$
used by \citet{RudermanSutherland1975}, especially for higher energy
photons.  Although this difference is only a factor of a few, as we
will point out later, the cascade efficiency in each generation
depends on the corresponding value of $\chiA$, and in a strong cascade
with several generations, the estimate for the final multiplicity will
be substantially affected by the value of $\chiA$.

Erber's expression is not applicable for pairs produced at low Landau
levels as it overestimates the opacity and, formally, solutions for
$\chiA$ obtained from eq.~(\ref{eq:chiA}) allows pair creation even
for $\epsilonG\sin\psi<2$.  In our analytical treatment we introduce a
limit on $\chiA$ to correct for the kinematic threshold.  For photons
above kinematic threshold from eq.~(\ref{eq:chi_def}) it follows that
\begin{equation}
  \label{eq:chiA_limit_derivation}
  \chiA>b\,,
\end{equation}
In all our analytical calculation we get $\SUB{\tilde{\chi}}{a}$ from
eq.~(\ref{eq:chiA}) and then use the maximum value of
$\SUB{\tilde{\chi}}{a}$ and $b$:
\begin{equation}
  \label{eq:chiA_limited}
  \chiA=\max\left( \SUB{\tilde{\chi}}{a},\;
    b\right)\,.
\end{equation}
In Fig.~\ref{fig:inverse_chi_corrected_rho6} we plot contours of
$1/\chiA$ which incorporate corrections to $\chiA$ due to the
kinematic threshold according to eq.~(\ref{eq:chiA_limited}). It is
clear from the plot that this correction affects only cases with high
magnetic field and low particle energies.

\subsection{Energy of escaping photons}
\label{sec:e_esc}

\begin{figure}[t]
  \centering
  \includegraphics[clip,width=\columnwidth]{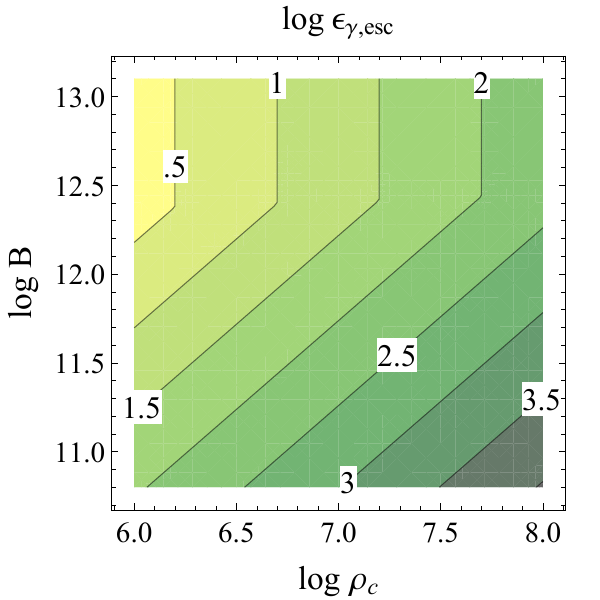}
  \caption{Energy of escaping photons: contours of
    $\log{\epsilonGesc}$ as a function of logarithms of the radius of
    curvature of magnetic field lines $\rhoC$ in cm and magnetic field
    strength $B$ in Gauss for $\ESC{s}=1$. }
  \label{fig:Eesc}
\end{figure}

As discussed above, photons escaping the cascade are those whose mfp
$l_\gamma$ is larger than the distance of significant magnetic field
attenuation $L_B$.  The formal criteria we use for calculating the
energy of escaping photons $\epsilonGesc$ is
$l_\gamma(\epsilonGesc)=\ESC{s} \RNS$; $\ESC{s}$ is a dimensionless
parameter quantifying the escaping distance in units of $\RNS$.  The
photon mfp is $l_\gamma{}=\rhoC\psiA$ and expressing the angle between
the photon momentum and the magnetic field at the point of absorption
$\psiA$ through $\chiA$ we get a (non-linear) equation for
$\epsilonGesc$
\begin{equation}
  \label{eq:eps_esc_eq}
  \epsilonGesc = 2\,\frac{\rhoC}{\ESC{s}\RNS}\frac{\chiA(\epsilonGesc,b,\rhoC)}{b}\,.
\end{equation}
The non-linearity in this equation is due to dependence of $\chiA$ on
$\epsilonGesc$, $b$, and $\rhoC$.  Using an interpolation table for
$1/\chiA$ this equation is very easy to solve numerically for all
reasonable values of physical parameters involved.

In Fig.~\ref{fig:Eesc} we plot energy of escaping photons,
$\log{\epsilonGesc}$ as a function of the radius of curvature of
magnetic field lines $\rhoC$ and magnetic field strength $B$ for
$\ESC{s}=1$.  For smaller values of $\ESC{s}$ the whole plot would
move to the left.  This figure shows (an obvious) trend that for
higher magnetic field and smaller radii of curvature, the energy of
escaping photons is lower, which allows for more cascade generations
and larger multiplicity. The break in contour lines around
$\log{B}\simeq12.4$ is due to the kinematic threshold.

For a given geometry of the magnetic field a more accurate analytical
expression for the energy of escaping photons can be derived using an
exact expression for $\psi$ along photon's travel path by expanding
the integral in eq.~(\ref{eq:tau_general}) around its upper endpoint,
as was done by \citet{Hibschman/Arons:pair_production::2001}.  However
here we try to explore different possible magnetic field configuration
-- exploring parameter space in $\rhoC$ -- and our simple estimate is
accurate to a factor of a few.

\section{Synchrotron Cascade}
\label{sec:synchrotron-cascade}

In this section we discuss the synchrotron cascade, where most of the
electron-positron pairs are created.  The synchrotron cascade is the
part of the whole cascade that is initiated by generation~0 photons.
In the synchrotron cascade each generation's primary photon is divided
into many (lower energy) next generation's pair-producing photons by
synchrotron radiation of freshly created pairs.

\subsection{Fraction of parent photon energy remaining \\ in the cascade}
\label{sec:frac-sync}

A high energy photon, when absorbed in the magnetic field, produces an
electron and a positron; the total energy of these particles is equal
to the energy of the photon.  Particle momenta just after production
have pitch angles equal to $\psiA$, when pair production takes place
well above threshold.  When pairs are created at high Landau levels,
as is the case in strong cascades, relativistic particles have
non-zero pitch angles and they radiate their perpendicular energy via
synchrotron radiation; in superstrong magnetic fields, this happens
almost instantaneously.  The component of particle momentum parallel
to the magnetic field is unaffected by synchrotron radiation and so
the final Lorentz factor of the particle
$\SUBB{\epsilon}{\pm}{\textsc{f}}$ will be
\begin{equation}
  \label{eq:gamma_F}
  \SUBB{\epsilon}{\pm}{\textsc{f}} =
  (1-\beta_{\parallel}^2)^{-1/2}\approx
  \SUBB{\epsilon}{\pm}{\textsc{i}}\left[ 1+(\psiA\,\SUBB{\epsilon}{\pm}{\textsc{i}})^2\right]^{-1/2}
\end{equation}
where $\beta_{\parallel}\equiv{}v_\parallel/c$ is particle velocity
along the magnetic field line and $\SUBB{\epsilon}{\pm}{\textsc{i}}$
is the initial Lorentz factor of the particle right after creation.
If the photon absorption happens at $\chiA<1$ -- which is indeed the
case near pulsar polar caps, see \S\ref{sec:opacity-gamma-b} -- with a
high degree of accuracy we can assume that the energy of the photon is
equally divided between the electron and positron
\citep[e.g.][]{DaughertyHarding1983}.  Expressing $\psiA$ in
eq.~(\ref{eq:gamma_F}) through $\chiA$, $\psiA=2\chiA/\epsilonG{}b$,
and the initial particle energy through the photon energy
$\epsilonG{}$, $\SUBB{\epsilon}{\pm}{\textsc{i}}=\epsilonG/2$ we get
$\SUBB{\epsilon}{\pm}{\textsc{f}}$ as a function of $\chiA$ and $b$
\begin{equation}
  \label{eq:gamma_F_chi}
  \SUBB{\epsilon}{\pm}{\textsc{f}} = \frac{\epsilonG}{2}\left[ 1+\left(\frac{\chiA}{b}\right)^2\right]^{-1/2}.
\end{equation}

\begin{figure}[t]
  \centering
  \includegraphics[clip,width=\columnwidth]{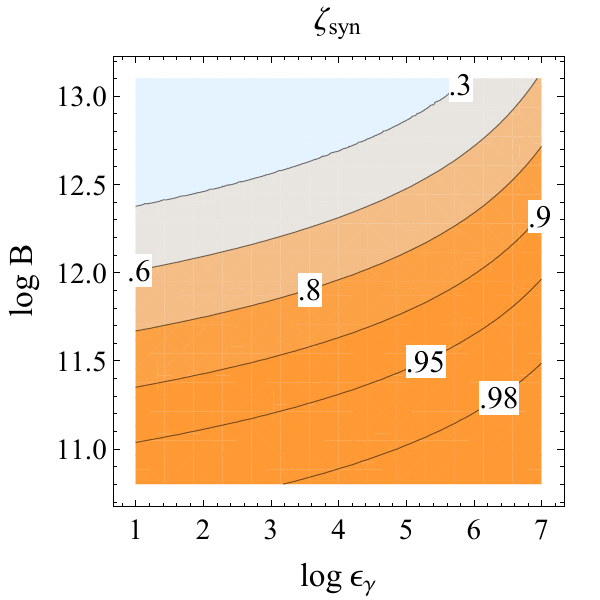}
  \caption{Fraction of the parent photon energy radiated as
    synchrotron photons by freshly created pairs: contours of
    $\SYN{\zeta}$ as a function of logarithms of the parent photon
    energy $\epsilonG$ and magnetic field strength $B$ in Gauss for
    $\rhoC=10^7\mbox{cm}$.}
  \label{fig:fe2SYN}
\end{figure}

The fraction $\SYN{\zeta}$ of photon energy radiated as synchrotron
photons, which is going into subsequent pair creation, is
\begin{equation}
  \label{eq:zeta_syn}
  \SYN{\zeta} = \frac{2(\SUBB{\epsilon}{\pm}{\textsc{i}}-\SUBB{\epsilon}{\pm}{\textsc{f}})}{\epsilonG} = 
  1 - \left[ 1+\left(\frac{\chiA}{b}\right)^2\right]^{-1/2}\,.
\end{equation}
Because of the kinematic threshold (\ref{eq:chiA_limit_derivation})
the minimum value of this fraction is
$\SYN{\zeta}|_{\chiA=b}\simeq0.292$, i.e. formally%
\footnote{as we mentioned above, the physics of near-threshold pair
  formation is more complicated and our simplified treatment is less
  accurate in this regime.}
at least $\simeq30\%$ of absorbed photon energy will go into
synchrotron radiation of created pairs. In Fig.~\ref{fig:fe2SYN} we
plot the fraction of pair-producing photon energy radiated as
synchrotron photons by freshly created pairs given by
eq.~(\ref{eq:zeta_syn}).  Contours of $\SYN{\zeta}$ are plotted as
functions of the photon energy $\epsilonG$ and magnetic field strength
$B$, the radius of curvature was assumed to be $\rhoC=10^7\mbox{cm}$.
The dependence of $\SYN{\zeta}$ on $\rhoC$ (via $\chiA$) is very weak,
and Fig.~\ref{fig:fe2SYN} is a good representation of how
$\SYN{\zeta}$ depends on $\epsilonG$ and $B$ for any $\rhoC$ of
interest.

It is evident from Fig.~\ref{fig:fe2SYN} that for higher magnetic
field strengths, $B\gtrsim\sci{3}{12}\mbox{G}$, and lower energies of
parent photons, a progressively smaller fraction of the parent photon
energy goes into synchrotron photons; the rest remains in the kinetic
energy of the created pairs moving along magnetic field lines.  The
portion of the parent photon energy energy left in kinetic energy of
pairs does not go into production of next generation pairs but is
``lost'' from the synchrotron cascade%
\footnote{the kinetic energy of pairs might be tapped by RICS cascade
  branches, see \S~\ref{sec:reson-inverse-compt}}.  The reason for
this is that for higher magnetic field strengths pairs are created
when the photon has a smaller pitch angle $\psiA$, so that a smaller
fraction of the photon energy goes into perpendicular pair energy, and
hence, a smaller fraction of the photon energy is emitted and remains
in the cascade.

\subsection{Number of secondary photons}

\begin{figure}[t]
  \centering
  \includegraphics[clip,width=\columnwidth]{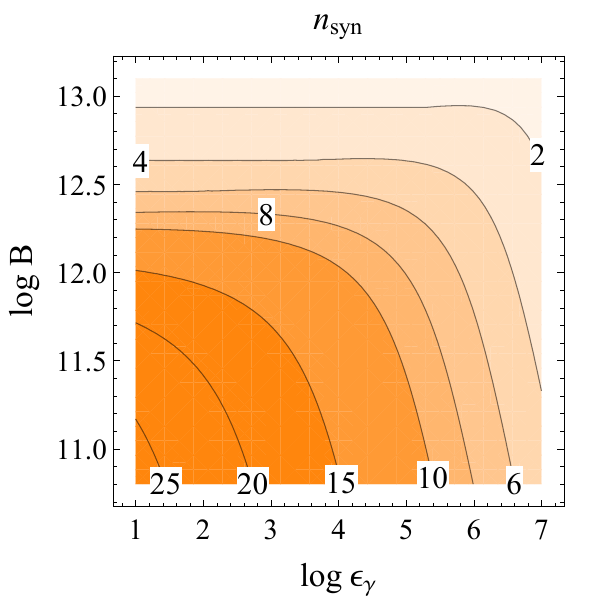}
  \caption{Number of synchrotron photons with the characteristic
    energy $\epsilonGSYN$ emitted at each
    $\gamma\rightarrow{}e^+e^-$ conversion: contours of $\SYN{n}$ as a
    function of logarithms of the parent photon energy $\epsilonG$ and
    magnetic field strength $B$ in Gauss for $\rhoC=10^7\mbox{cm}$.}
  \label{fig:n_sync}
\end{figure}

At each pair creation event the parent photon is effectively
transformed into an electron-positron pair and lower energy
synchrotron photons.  Those photons become parent photons for the next
cascade generation or escape the magnetosphere, terminating the
cascade.

The characteristic energy of synchrotron photons emitted by a newly
created particle in terms of quantities used in this paper is given by
\begin{equation}
  \label{eq:epsilon_sync}
  \epsilonGSYN = \frac{3}{2}b\,\psiA\SUBB{\epsilon}{\pm}{\textsc{i}}^2.
\end{equation}
The number of synchrotron photons with the characteristic energy
$\epsilonGSYN$ -- these photons carry most of the energy of
synchrotron radiation -- emitted at each event of conversion of a
parent photon with energy $\epsilonG$ into an electron-positron pair
is
\begin{equation}
  \label{eq:n_sync}
  \SYN{n} \simeq\frac{\SYN{\zeta}\,\epsilonG}{\epsilonGSYN} = \frac{4}{3}\frac{\SYN{\zeta}}{\chiA}.
\end{equation}
In eq.~(\ref{eq:n_sync}) both $\SYN{\zeta}$ and $\chiA$ are functions
of $B$, $\epsilonG$, and $\rhoC$.  In Fig.~\ref{fig:n_sync} we plot
contours of $\SYN{n}$ as functions of the photon energy and magnetic
field strength, the radius of curvature was assumed to be
$\rhoC=10^7\mbox{cm}$.  Two clear trends are visible on this plot: the
lower the energy of the primary photon the larger the number of
secondary synchrotron photons produced at each conversion event, and
the higher the magnetic field the smaller is the number of synchrotron
photons.  The first one is a general trend of emission processes when
higher energy particles emit less photons which, however, have larger
energies.  The second trend is due to the suppression of the
synchrotron cascade discussed above, in \S\ref{sec:frac-sync}.

\subsection{Multiplicity of synchrotron cascade}
\label{sec:mult-synchr-casc}

The generation $i+1$ cascade photon is a synchrotron photon emitted at
the event of pair creation by a photon of generation $i$.  Expressing
$b$, $\psiA$ and $\epsilonG^{(i)}$ through $\chiA$ and $\epsilonG$
from eq.~(\ref{eq:epsilon_sync}) we get for the characteristic energy
of the next generation photon
\begin{equation}
  \label{eq:eSYN-next-gen}
  \epsilonG^{(i+1)}=\frac{3}{4}\chiA \epsilonG^{(i)}\,,
\end{equation}
where $\epsilonG^{(i)}$ and $\epsilonG^{(i+1)}$ are energies of $i$'th
and $(i+1)$'th generation photons.  The photon energy degrades with
each successive generation of the cascade.  This degradation
accelerates as the cascade proceeds through generations because with
the decrease of the photon energy $\chiA$ increases, see
Fig.~\ref{fig:inverse_chi}.

The photon mfp in a constant magnetic field goes as
$l_\gamma\propto\chiA/\epsilonG$.  For photon energies
$\epsilonG\lesssim{}10^4$, when $\chiA\gtrsim{}10$, the mfp of
successive generations increases by at least an order of magnitude in
each successive generation; therefore the crudeness of our
approximation for estimating the energy of escaping photons,
eq.~(\ref{eq:eps_esc_eq}), could affect only the last cascade
generation.

The rapid energy degradation results in a rather small number of
generations in polar cap cascades as the energy of pair-producing
photons rapidly reaches the threshold energy.  On the other hand, the
number of emitted synchrotron photons increases with the decrease of
the energy of the parent photon, see Fig.~\ref{fig:n_sync};
consequently synchrotron cascades can have quite large multiplicities
despite a small number of generations.

Each cascade photon creates 2 particles at the moment of absorption,
and particles are produced until $\epsilonG^{(i+1)}\ge\epsilonGesc$.
The total number of particles generated in synchrotron cascades
initiated by a primary photon can be calculated by summation over all
generations of the number of pairs produced in each cascade
generation.  The algorithm for this calculation is shown in
Appendix~\ref{sec:algorithms}, algorithm~\ref{alg:sync_mult}.  We will
use this algorithm for calculation of the polar cap cascade
multiplicity after we discuss curvature radiation, the radiation
mechanism responsible for generating the primary photons for
synchrotron cascades in young energetic pulsars.

Finally we wish to point out how the magnetic field strength affects
the multiplicity of the synchrotron cascade.  From discussions
presented in \Ss\ref{sec:e_esc},~\ref{sec:frac-sync} it is evident
that, for the same energy of the primary photon, a higher magnetic
field strength results in (i) reducing the final energy of escaping
photons, therefore increasing cascade efficiency, and at the same time
(ii) in suppression of cascade efficiency by forcing a smaller
fraction of the photon energy to go into perpendicular energy of the
created pairs and so short-circuiting the cascade.  Therefore, there
should be a ``sweet spot'' in magnetic field strength where the
synchrotron cascade is most efficient.

\section{Curvature Radiation}
\label{sec:curvature-radiation}

In this section we discuss how primary particles emit photons which
``launch'' the synchrotron cascade.  As we discussed above the most
efficient process for supplying the primary (generation 0) photons in
young energetic pulsars is curvature radiation.

\subsection{Fraction of the primary particle energy going into the
  cascade}
\label{sec:feCR}

\begin{figure}[t]
  \centering
  \includegraphics[clip,width=\columnwidth]{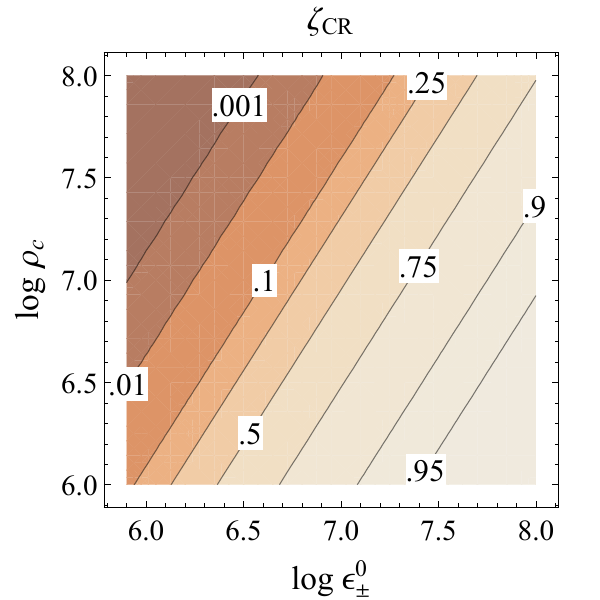}
  \caption{Fraction of the primary particle energy emitted as CR
    photons over the distance $\CR{s}=1$: contours of $\CR{\zeta}$ as
    a function of logarithms of the initial particle energy
    $\epsilonP^0$ and the radius of curvature of magnetic field lines
    $\rhoC$ in cm. }
  \label{fig:fe2CR}
\end{figure}

Ultrarelativistic particles moving along curved magnetic field lines
emit electromagnetic radiation with power \citep{Jackson1975}
\begin{equation}
  \label{eq:CR_energy_losses}
  \CR{P} = \frac{2}{3} \frac{e^2}{m_ec^5}\left(\frac{c^2}{\rhoC}\right)^2 \epsilonP^4\,,
\end{equation}
where $\CR{P}$ is normalized to $m_ec^2/\mbox{sec}$, $\epsilonP$ is
the particle's energy normalized to $m_ec^2$; we do not distinguish
between electrons and positrons.  Since we are treating the pair
generation problem separately from the problem of particle
acceleration, we consider cascades produced by particles injected in a
region with screened electric field, so that particles are not
accelerated and only lose energy to CR.  The particle's energy
decreases with time according to the equation of motion
\begin{equation}
  \label{eq:CR_es_equation}
  \frac{d\epsilonP}{dt} = - \CR{P}.
\end{equation}
Solving eq.~(\ref{eq:CR_es_equation}) we get for the particle energy
after it travels the distance $s$ from the injection point
\begin{equation}
  \label{eq:CR_es}
  \epsilonP(s) = \epsilonP^0 \left[1 + 3 H \frac{(\epsilonP^0)^3}{\rhoC^2} s\right]^{-1/3}
\end{equation}
\citep[see also][]{Harding1981}, where $s$ is normalized to $\RNS$,
$\epsilonP^0$ is the initial particle energy; constant $H$ is defined
as $H=(2/3)\RNS{}r_e\approx\sci{1.88}{-7}\mbox{cm}^2$, where
$r_e=e^2/m_ec^2$ is the classical electron radius.

The fraction \CR{\zeta} of the initial particle energy lost to
curvature radiation after a particle has traveled distance $\CR{s}$ is
\begin{equation}
  \label{eq:CR_zeta_CR}
  \CR{\zeta}(\CR{s}) = 1- \frac{\epsilonP(\CR{s})}{\epsilonP^0}.
\end{equation}
If the energy of these CR photons goes into creation of
electron-positron pairs, $\CR{\zeta}$ gives the efficiency of the CR
part of the full cascade.  The electric field in the acceleration zone
transforms electromagnetic energy into particle's kinetic energy,
which is then radiated as pair producing photons.  Only the photon's
energy can be divided in chunks carried by a large number of pairs.
The cascade will have high efficiency if (i) primary particles have
high energy, (ii) emit most of their energy as photons, and (iii)
inject these photons in the region where the synchrotron cascade can
work effectively, i.e. in a region close to the NS which is smaller
than the characteristic scale of magnetic field variation $L_B$.

In Fig.~\ref{fig:fe2CR} we plot the fraction of the primary particle
energy emitted as CR photons after the particle has traveled distance
$\CR{s}=1$.  Shown are contours of $\CR{\zeta}(\CR{s})$ as a function
of the initial particle energy $\epsilonP^0$ and the radius of
curvature of magnetic field lines $\rhoC$.  For smaller values of
$\CR{s}$ the whole plot would move to the right.  CR is most efficient
in transferring particle energy into the cascade in the parameter
space corresponding to the lower right triangular region of
Fig.~\ref{fig:fe2CR}.  Going from the upper left (smaller
$\epsilonP^0$, larger $\rhoC$) to the lower right (larger
$\epsilonP^0$, smaller $\rhoC$) on this plot, not only the particle
energy increases but also the fraction of the energy which can go into
the cascade.

For a certain range of $\epsilonP^0$ and $\rhoC$ the energy put into the cascade
by the primary particle grows faster than the energy of that particle, i.e. the
fraction of particle's energy going into the cascade increases stronger than
linearly with the energy of the particle.  For more or less regular global
magnetic fields, with $\rhoC\gtrsim10^7\mbox{cm}$, the transition between
effective and ineffective CR cascades occurs at particle energies
$\epsilonP\sim10^7$, and the efficiency is very sensitive to the particle
energy.  In this parameter range even a modest increase of the primary particle
energy can result in a large boost of the cascade multiplicity.

\subsection{Energy of CR photons and critical particle energy}
\label{sec:eCR-eThreshold}

\begin{figure}[t]
  \centering
  \includegraphics[clip,width=\columnwidth]{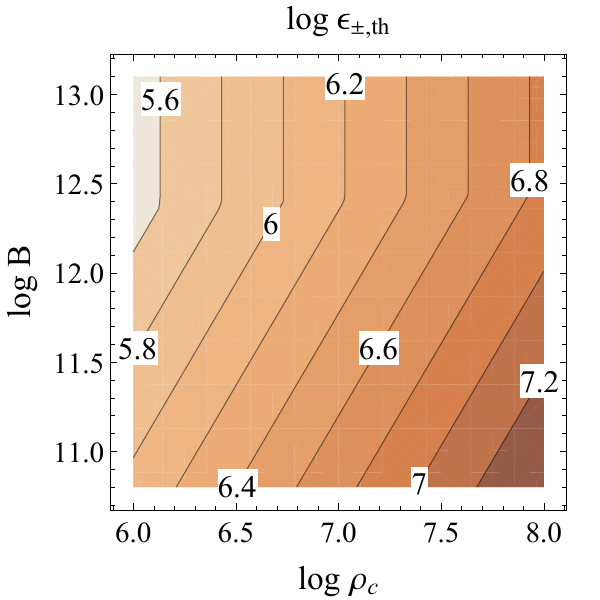}
  \caption{Critical particle energy above which it can emit
    pair-producing photons via curvature radiation: contours of
    $\log\SUBB{\epsilon}{\pm}{th}$ as a function of logarithms of the
    radius of curvature of magnetic field lines $\rhoC$ in cm and
    magnetic field strength $B$ in Gauss for $\ESC{s}=1$.}
  \label{fig:eP_thresh}
\end{figure}

The characteristic energy of CR photons emitted by a particle with the
energy $\epsilonP$ is \citep{Jackson1975}
\begin{equation}
  \label{eq:CR_ephot}
  \epsilonGCR = \frac{3}{2} \frac{\lambdaC}{\rhoC} \epsilonP{}^3
  \approx\sci{5.8}{3}\,\SUB{\rho}{c,\,7}^{-1}\:\SUBB{\epsilon}{\pm}{7}^3\,,
\end{equation}
where $\SUB{\rho}{c,\,7}\equiv{}\rhoC/10^7\mbox{cm}$ and
$\SUBB{\epsilon}{\pm}{7}\equiv\epsilonP/10^7$.  The number of CR
photons emitted by the particle while traveling distance $ds$
normalized to $\RNS$ is
\begin{equation}
  \label{eq:n_CR}
  \frac{d\,\CR{n}}{ds} \simeq \frac{\RNS}{c} \frac{\CR{P}}{\epsilonGCR}
\end{equation}
Each CR photon above the pair formation threshold will be a primary
photon for the synchrotron cascade discussed in
\S\ref{sec:synchrotron-cascade}.  The critical energy at which primary
particles can produce pair-creating CR photons can be calculated by
equating $\epsilonGCR$ given by eq.~(\ref{eq:CR_ephot}) to the escape
photon energy $\epsilonGesc$ from eq.~(\ref{eq:eps_esc_eq}).  In
Fig.~\ref{fig:eP_thresh} we plot the critical particle energy which
could initiate pair production with CR photons
$\SUBB{\epsilon}{\pm}{th}$ for $\ESC{s}=1$.  Shown are contours of
$\log\SUBB{\epsilon}{\pm}{th}$ as a function of the radius of
curvature of magnetic field lines $\rhoC$ and magnetic field strength
$B$.  Primary particles should have energies $\epsilonP\gtrsim{}10^6$
to be able to initiate pair production via CR.

\subsection{Multiplicity of CR-synchrotron cascade}
\label{sec:mult-CR-synch}

\begin{figure*}
  \includegraphics[clip,width=\textwidth]{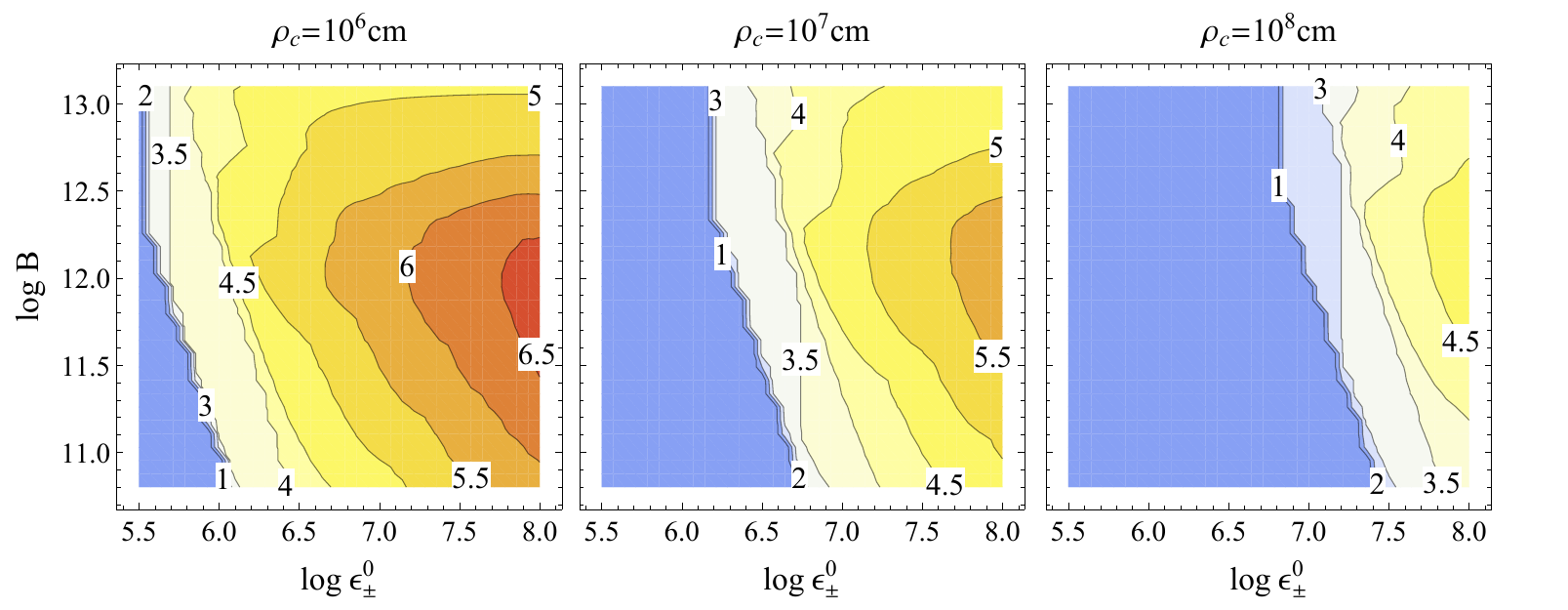}
  \caption{Multiplicity of CR-synchrotron cascade: contours of
    $\log\CRsyn{\kappa}$ as a function of logarithms of the primary
    particle energy $\epsilonP^0$ and magnetic field strength $B$ in
    Gauss for three values of the radius of curvature of magnetic
    field lines $\rhoC=10^6, 10^7, 10^8\mbox{cm}$.  Assumed values for
    characteristic lengths: $\CR{s}=\ESC{s}=1$.}
  \label{fig:multiplicity}
\end{figure*}

\begin{figure*}
  \includegraphics[clip,width=\textwidth]{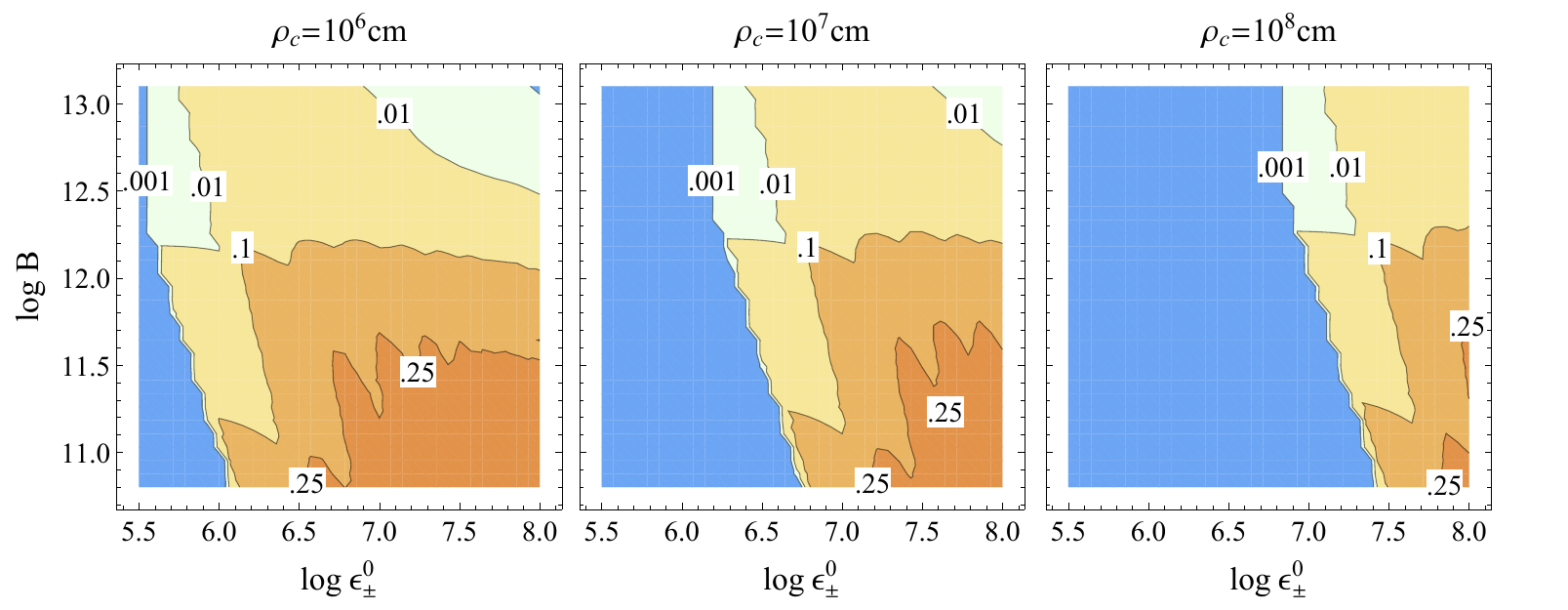}
  \caption{Efficiency of CR-synchrotron cascade as given by
    eq.~(\ref{eq:zeta-cascade}): contours of $\CRsyn{\zeta}$ as a
    function of logarithms of the primary particle energy
    $\epsilonP^0$ and magnetic field strength $B$ in Gauss for three
    values of the radius of curvature of magnetic field lines
    $\rhoC=10^6, 10^7, 10^8\mbox{cm}$. Assumed values for
    characteristic lengths: $\CR{s}=\ESC{s}=1$.}
  \label{fig:efficiency}
\end{figure*}

The multiplicity of the the CR-synchrotron cascade -- the total number
of particles produced by a single primary electron or positron
accelerated in the gap -- can be computed by multiplying the number of
CR photons $\CR{n}$, eq.~(\ref{eq:n_CR}), by the number of particles
produced in the synchrotron cascade initiated by these photons
$\SYN{n}$, eq.~(\ref{eq:n_sync}), and integrating it over the distance
where CR can initiate a cascade
\begin{equation}
  \label{eq:kappa-CR-Sync}
  \CRsyn{\kappa} = \int_0^{\CR{s}}\SYN{n}\frac{d\,\CR{n}}{ds}\,ds\,. 
\end{equation}
The actual algorithm we use to compute the total multiplicity is the
Algorithm~\ref{alg:total_mult} from Appendix~\ref{sec:algorithms}. Integration
in eq.~(\ref{eq:kappa-CR-Sync}) is done assuming constant values for $B$ and
$\rhoC$, as discussed in \S~\ref{sec:opacity-gamma-b}.  In
Fig.~\ref{fig:multiplicity} we plot $\CRsyn{\kappa}$ as a function of the
initial particle energy $\log\epsilonP^0$ and magnetic field strength $\log{}B$
for three different radii of curvature of magnetic field lines
$\rhoC=10^6,10^7,10^8$~cm assuming $\CR{s}=\ESC{s}=1$.

It is evident from these plots that in a dipolar magnetic field, with
$\rhoC\approx10^8\mbox{cm}$, the maximum achievable multiplicity is
$\CRsyn{\kappa}\sim\sci{\mbox{few}}{4}$ even in cascades initiated by
extremely energetic primary particles.  If the radius of curvature is
an order of magnitude less, a rather high multiplicity
$\CRsyn{\kappa}\gtrsim10^5$ could be achieved in polar cap cascades
for magnetic field strength $B\sim10^{12}$~G and particle energies
$\epsilonP\gtrsim10^7$, parameters quite realistic for young pulsars.
For strongly non-dipolar magnetic field, with
$\rhoC\approx10^6\mbox{cm}$ the multiplicity can be another order of
magnitude higher $\CRsyn{\kappa}\sim\sci{\mbox{few}}{6}$.

The properties of CR do not depend on the strength of the magnetic
field, therefore the effect of the magnetic field strength on the
multiplicity of CR-synchrotron cascade is due to the synchrotron
cascade and how many CR photons pair produce.  As discussed at the end
of \S~\ref{sec:mult-synchr-casc} there should be an optimum magnetic
field strength where the multiplicity is the highest; the multiplicity
decreases both for higher (due to larger energy left in particle
motion along magnetic field lines) as well as lower magnetic field
(due to increase of the energy of escaping photons).  This trend is
clearly visible on all plots of Fig.~\ref{fig:multiplicity}.  The
highest cascade multiplicity is achieved for magnetic field around
$B\sim10^{12}$~G.  The value of this optimum magnetic field grows
slightly with increase of $\rhoC$, but it stays around
$\sim\sci{\mbox{few}}{12}$~G even for dipolar magnetic fields.  This
is noteworthy in view of the fact that $B\sim10^{12}$~G is the typical
value of magnetic field strength for normal pulsars.  For any given
energy of the primary particle the decrease of the cascade
multiplicity towards stronger magnetic fields is faster than for
weaker fields.

The dependence of cascade multiplicity on the initial energy of the primary
particle is non-linear.  Let us consider what happens when the initial particle
energy $\epsilonP^0$ goes from the highest to the lowest value (horizontal
direction in plots of Fig.~\ref{fig:multiplicity}).  For the highest values of
$\epsilonP^0$ multiplicity decreases uniformly, but then it drops by an order of
magnitude in a rather small range of $\epsilonP^0$ (for $B\sim10^{12}$G it
happens around $\epsilonP^0\sim10^{6.3}$ for $\rhoC=10^6$cm,
$\epsilonP^0\sim10^{6.8}$ for $\rhoC=10^7$cm, and $\epsilonP^0\sim10^{7.4}$ for
$\rhoC=10^8$cm).  After about half a decade of $\epsilonP^0$ values the
multiplicity drops again to 1, when no particles can be produced (for
$B\sim10^{12}$G it happens around $\epsilonP^0\sim10^{5.7}$ for $\rhoC=10^6$cm,
$\epsilonP^0\sim10^{6.3}$ for $\rhoC=10^7$cm, and $\epsilonP^0\sim10^{7}$ for
$\rhoC=10^8$cm).  The first effect is due to the decrease of the efficiency of
CR, discussed in \S\ref{sec:feCR}.  For lower initial energies of the primary
particle there is less energy available to create pairs, not only because the
particle energy is smaller, but also due to smaller efficiency of CR in
producing photons -- the primary particle keeps most of its energy, depositing
only a small fraction of it in the cascade zone.  The drop in the efficiency of
CR $\CR{\zeta}$, where it becomes less than 10\% (see Fig.~\ref{fig:fe2CR})
manifests in a rapid decrease of cascade multiplicity -- by an order of
magnitude -- on all plots of Fig.~\ref{fig:multiplicity} for magnetic field
strengths where the maximum multiplicity is achieved.  This drop in
$\CRsyn{\kappa}$ is most prominent for $B\sim10^{12}$G and is less pronounced
for both higher and lower magnetic field strengths due to lower efficiency of
the cascade discussed above.  The second drop in $\CRsyn{\kappa}$, towards 1, is
due to the threshold in pair formation -- for those particle energies CR photons
have too low an energy to initiate a cascade.  The blue region on the plots of
Fig.~\ref{fig:multiplicity} show the parameter space where no particles can be
produced by the CR-synchrotron cascade.  It does not mean, however, that no
pairs can be produced in the polar cap cascades for such primary particle
energies.  Instead of CR, the primary photons for synchrotron cascade will be
produced by inverse Compton scattering of thermal photons emitted by the NS,
however, those primary photons will have much lower energies and multiplicities
of such cascades will be quite low
\citep[see][]{Harding/Muslimov:heating_2::2002}.

We find it also instructive to compare the multiplicity of the
CR-synchrotron cascade with the theoretical upper limit on cascade
multiplicity $\SUB{\kappa}{max}$, given by eq.~(\ref{eq:kappa_max}) in
\S\ref{sec:cascades-overview}. The ratio
\begin{equation}
  \label{eq:zeta-cascade}
  \CRsyn{\zeta}=\frac{\CRsyn{\kappa}}{\SUB{\kappa}{max}}
\end{equation}
can be considered as the efficiency of splitting the energy of the
primary particle into pairs.  In Fig.~\ref{fig:efficiency} we plot
$\CRsyn{\zeta}$ for the same values of parameters as $\CRsyn{\kappa}$
in Fig.~\ref{fig:multiplicity}.  Despite lower multiplicity for
smaller values of $B$, the cascade efficiency is higher, i.e. more of
the initial energy of the primary particle goes into pair formation as
opposed to the energy of escaping photons and kinetic energy of pairs
and primary particles.  This trend is discussed in
\S\ref{sec:frac-sync},~\ref{sec:feCR}.  It is interesting to note that
for $B\lesssim\sci{\mbox{few}}{11}$~G the cascade can be quite
efficient in splitting a noticeable fraction of primary particle
energy into pairs.  For magnetic field $\sim10^{12}$~G the fraction of
the primary particle's energy going into pair production saturates at
$\sim10\%$; this is the limiting efficiency of the highest possible
multiplicity cascade in a typical pulsar.  The dependence of
$\CRsyn{\zeta}$ on $\epsilonP^0$ is similar to the dependence of
$\CRsyn{\kappa}$ on $\epsilonP^0$.

\section{Particle acceleration}
\label{sec:particle-acceleration}

\subsection{Overview of particle acceleration regimes}
\label{sec:overview-particle-acceleration}

\begin{figure*}
  \includegraphics[clip,width=\textwidth]{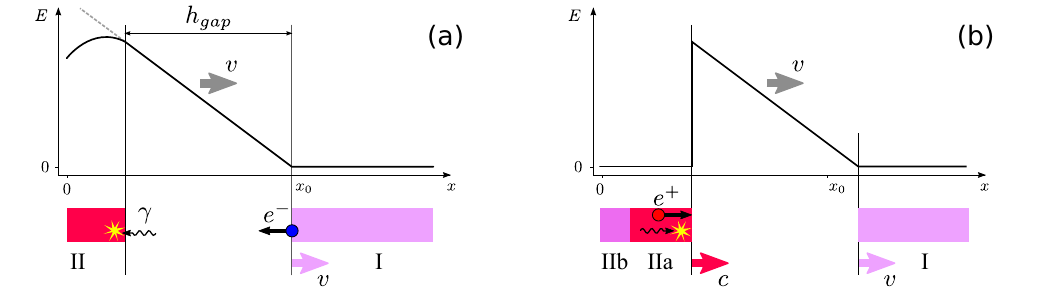}
  \caption{Schematic representation of gap formation and evolution for
    cascades in Ruderman-Sutherland regime with $\jm/\GJ{j}>0$. See text
    for explanation.}
  \label{fig:accelerator}
\end{figure*}

In this section we will get an estimate for the energy of primary
cascade particles.  In the following discussion we will rely on
results of self-consistent modeling of pair cascades by
\citet{Timokhin2010::TDC_MNRAS_I} [T10] and \citet{TimokhinArons2013}
[TA13].  First, we give a brief overview of how particle acceleration
proceeds according to these simulations.

Whether and how the pair formation along given magnetic field lines
occurs depends on the ratio $\jm/\GJ{j}$ of the current density
required to support the twist of magnetic field lines in the pulsar
magnetosphere \citep[e.g.][]{Timokhin2006:MNRAS1,BaiSpitkovsky2010a},
$\jm\equiv(c/4\pi)\,|\nabla\times\mathbf{B}|$, to the local GJ current
density, $\GJ{j}\equiv\GJ{\eta}c$, where $\GJ{\eta}=B/Pc$ is the GJ
charge density.

For the \citet{RudermanSutherland1975} cascade model, where particles
cannot be extracted from the NS surface, effective particle
acceleration and pair formation is possible for almost all values of
$\jm/\GJ{j}$ (T10).  In the space charge limited flow regime, first
discussed by \citet{AronsScharlemann1979}, pair formation is not
possible if $0<\jm/\GJ{j}<1$, but is possible for all other values of
$\jm/\GJ{j}$ (TA13).  Pair formation is always non-stationary: an
active phase when particles are accelerated to ultrarelativistic
energies and give rise to electron-positron cascades -- with a burst
of pair formation -- is followed by a quiet phase when recently
generated dense electron-positron plasma screens the electric field
everywhere.

As pair plasma leaves the active region, it flows into the
magnetosphere, and later into the pulsar wind.  When the density of
the pair plasma drops below the minimum density necessary for
screening of the electric field a gap appears -- a charge starved
region where the electric field is very strong, of the order of the
vacuum electric field.  To screen the electric field, the plasma
density must be high enough to provide \emph{both} the GJ charge
density $\GJ{\eta}$ and the imposed current density $\jm$.  The
transition between the region(s) still filled with plasma (hereafter
we call it ``plasma tail'') and the gap is sharp -- plasma still
capable of screening the electric field moves in bulk, its density is
close to the critical density.  The plasma density drops abruptly at
the gap boundary; within the gap the particle number density is much
smaller than the critical density
(cf. Fig.~\ref{fig:xp_xpdensity_RS},~\ref{fig:xp_xpdensity_superGJ} in
\S~\ref{sec:flux-primary}).  The motion of the boundary between the
plasma filled region and the gap sets the gap's growth rate.  Some
particles enter the gap and are accelerated to ultrarelativistic
energies.  The gap grows until the energy of particles accelerated
there becomes sufficient to start electron-positron cascades and the
cycle repeats.  The gap is not stationary; in almost all cases it
moves as a whole -- after its growth has been terminated by newly
created pairs its upper and low boundaries are moving in the same
direction.  The gap can move, keeping its size for a long time, or it
might disappear rather quickly.  The details of gap behavior -- where
the gap appears, what direction it moves, how fast it disappears --
depend on the ratio $\jm/\GJ{j}$.  However, despite these differences,
the way in which the highest energy particles are accelerated is very
similar in any regime which allows pair formation studied in T10,
TA13.  Namely, the size of the charge starved region grows as the tail
of pair plasma moves, the bulk velocity of the tail $v$ sets the rate
of the gap expansion. Particles entering the gap from the tail are
accelerated in a larger and larger gap, until they are able to produce
pair-producing photons.  The place where these photons are absorbed
and produce pairs is the other boundary of the gap.

\subsection{Energy of primary particles}
\label{sec:energy-primary}

In this section we obtain a quantitative estimate for the maximum
energy of accelerated particles using as an example the case of the
Ruderman-Sutherland (RS) cascade, when particles cannot be supplied
from the surface of the NS. As we mentioned before, gap formation and
particle acceleration in the space charge limited flow regime, when
pair creation is allowed, is very similar to the RS case and estimates
for particle energies obtained in this section are applicable for the
space charge limited flow with $\jm/\GJ{j}>1$ and $\jm/\GJ{j}<0$.  The
GJ charge density is positive and we consider the case when the ratio
$\jm/\GJ{j}>0$.  In Fig.~\ref{fig:accelerator} we show a schematic
picture of how particles are accelerated in this cascade.  On the top
of each figure we show the electric field in the accelerating region
and on the bottom a schematic representation of plasma motion in and
around the gap; plot (a) corresponds to the time when the electric
field screening has just started, plot (b) shows a well-developed gap
moving into the magnetosphere. These schematic plots illustrate
results of actual simulations of RS cascades shown in Fig.~3,4,11 and
13 in T10.

At the beginning of the burst of pair formation, the gap appears at the NS
surface and its upper boundary is the ``tail'' of plasma left from the previous
burst of pair formation, where the particle number density is still high enough
to screen the electric field (region I in Fig.~\ref{fig:accelerator}).
Electrons and positrons in this tail are trapped in electrostatic oscillations
and the bulk velocity of this tail $v$ is sub-relativistic, but for large
current densities (around or greater than $\GJ{j}$) it is quite close to $c$.
Electrons from this tail which get to the gap boundary are pulled into the gap
and are accelerated toward the NS.  As the tail moves, the gap grows; the
current and charge density in the gap is due to the flux of electrons from the
tail and so it remains constant within the gap.  The gap growth is stopped when
electrons reach an energy high enough to produce pair-creating photons. This
first-generation of pairs start screening the electric field -- electrons move
toward the NS and positrons are accelerated toward the magnetosphere and start
producing pair creating photons as well (region II in
Fig.~\ref{fig:accelerator}(a)). In numerical simulations (T10) the
first-generation positrons moving toward the magnetosphere have approximately
the same energies as the primary electrons which initiated the discharge.
Because those positrons are ultrarelativistic they practically co-move with the
photons and so new pairs are injected close to their parent particles making a
blob of pair plasma moving into the magnetosphere (region IIa in
Fig.~\ref{fig:accelerator}(b))%
\footnote{see also Fig.~\ref{fig:xp_xpdensity_RS} in \S~\ref{sec:flux-primary}
  where we show a snapshot from numerical simulations of the cascade
  corresponding to the stage shown in Fig.~\ref{fig:accelerator}(b)}.
This blob is the lower boundary of the accelerating gap, and the gap
exists until this blob catches up with the tail from the previous pair
formation cycle. For large current densities this can take a while as
$v$ is close to $c$. Plasma leaking from the blob forms the new tail
(region IIb on Fig.~\ref{fig:accelerator}(b)).

In the discharge described above primary particles are moving in both directions
and initiate cascades toward the NS (electrons) and the magnetosphere
(positrons).  As the discharges happen close to the NS surface, the cascade can
fully develop only in the direction of the magnetosphere -- particles moving
toward NS slam onto the star's surface before they can produce a lot of pairs.
For RS discharges the primary, generation 0, particles initiating the full
cascade in Fig.~\ref{fig:full_cascade} are positrons in region IIa in
Fig.~\ref{fig:accelerator}(b).  As we mentioned above, the energy of those
positrons is very close to the energy of the primary electrons and here, for the
sake of simplicity, we provide estimates only for the energy of primary
electrons%
\footnote{The reason for both kinds of primary particles, electrons and
  positrons, acquiring almost the same energies is that the potential drop
  experienced by each of them is regulated by the process of pair formation,
  rather than by the details of their acceleration.  We did analytical estimates
  for the final energies of the first-generation positron based on the model
  presented in this section; the difference between energies of the primary
  electrons and first-generation positrons in the frame of the model is about
  2\%.}.

The evolution of the electric field in any given point $x$ and moment
of time $t$ is given by (see e.g. eq.~1 in TA13)
\begin{equation}
  \label{eq:E_jjm}
  \frac{\partial{E}}{\partial{t}}(x,t) = 
  -4\pi\left( j(x,t) - \jm\right )
  \equiv -4\pi\tilde{\jmath}\,,
\end{equation}
$j$ is the actual current density along a given magnetic field line
and $\jm$ is the current density imposed by the magnetosphere.  The
difference $\tilde{\jmath}\equiv{}j - \jm$ in the gap remains
constant.  When the upper boundary of the gap moves with the constant
speed $v$ this equation can be integrated to get the electric field in
the gap
\begin{equation}
  \label{eq:E_x_t}
  E(x,t) = E(x_0,t_0) + 4\pi \tilde{\jmath}\, \frac{x_0-x}{v} + 4\pi \tilde{\jmath}\, (t-t_0)
\end{equation}
Where $E(x_0,t_0)$ is the electric field within the gap at the moment
$t_0$ at the point $x_0$.  If we assume that the gap boundary is at
$x_0$ at the moment $t_0$, then $E(x_0,t_0)=0$.

Electrons enter the gap from above and are quickly accelerated by the
strong electric field and move with relativistic speed practically
from the moment they leave the plasma tail.  If a particle enters the
gap at $t_0$ (in point $x_0$) its coordinate is $x_\pm=x_0-c(t-t_0)$,
substituting $x_\pm$ into eq.~(\ref{eq:E_x_t}) the electric field seen
by that particle is given by
\begin{equation}
  \label{eq:E_xp}
  E(x_p) = \frac{4\pi}{c} \tilde{\jmath}\, \left( 1+\frac{c}{v}
  \right) \left( x_0 - x_\pm \right)
  \equiv 4\pi \GJ{\eta}^0\, \xi_j\, \lP
\end{equation}
In the last step in eq.~(\ref{eq:E_xp}) we denote with
$\lP\equiv|x_\pm-x_0|$ the distance traveled by the particle in the gap 
and introduce $\xi_j$ defined as
\begin{equation}
  \label{eq:xi_j_def}
  \xi_j\equiv \frac{\tilde{\jmath}}{\GJ{j}^0} \left( 1+\frac{c}{v} \right)\,,
\end{equation}
where $\GJ{j}^0$ is the GJ current density in an aligned rotator
\begin{equation}
  \label{eq:jGJ_0}
  \GJ{j}^0\equiv\GJ{\eta}^0c=\frac{B}{P}\,.
\end{equation}
$\xi_j$ is a factor which shows how stronger/weaker the electric field
in the gap is compared to the situation of a static vacuum gap in an
(anti-)aligned rotator, like the one considered by
\citet{RudermanSutherland1975}.  $v\simeq\mbox{const}$ is a good
approximation to the numerical results and so
$\xi_j\simeq\mbox{const}$.  In cascades along magnetic field lines
where $\jm$ is close to the local value of $\GJ{j}$ in an aligned
rotator $\xi_j\sim2$, for the same situation in a pulsar with
inclination angle of $60^{\circ}$, $\xi_j\sim1$.

For energy losses dominated by curvature radiation, free acceleration
is a good approximation for $B\gtrsim{}10^{11}$~G (see
Appendix~\ref{sec:trans-radi-react}).  If radiation losses are
negligible, the particle's equation of motion is
\begin{equation}
  \label{eq:dpdt}
  \frac{d p}{d t}=-eE\,,
\end{equation}
where $p=mc\gamma$ is particle's momentum.  In terms of the distance
traveled by the particle in the gap $\lP=c(t-t_0)$ with $E$ given by
eq.~(\ref{eq:E_xp}), particle equation of motion can be written as
\begin{equation}
  \label{eq:dpds}
  \frac{d p}{d \lP}=\frac{4\pi e}{c} \GJ{\eta}^0 \xi_j \,\lP\,.
\end{equation}
Integrating this equation and expressing $\GJ{j}$ through pulsar
parameters we get for particle energy
\begin{equation}
  \label{eq:epsilonP_s}
  \epsilon_\pm = \frac{2 \pi}{B_q \lambdaC c}\xi_j \frac{B}{P}\, \lP^2
\end{equation}

The distance the primary particle travels in the region of unscreened
electric field $\GAP{l}$ -- the size of the gap as seen by the moving
particle -- is the sum of the distance the particle travels before
emitting pair-producing photons terminating the gap $\lPeGAP$ and the
distance these protons travel until the absorption point
$\GAParg{l}{\gamma}$
\begin{equation}
  \label{eq:l_gap_def}
  \GAP{l}=\lPeGAP+\GAParg{l}{\gamma}\,.
\end{equation}
For any given particle, the larger the distance $\lPe$ the particle
travels to the emission point, the higher the particle energy and the
energy of CR photons it emits, and so the smaller is the distance
traveled by the photon until the absorption point $l_\gamma$.  The
distance the particle travels in the gap $\GAP{l}$ is the minimum
value of $l=\lPe+l_\gamma$ because once the first pairs are injected
the avalanche of pair creation will lead to screening of the electric
field.  The photon mean free path $l_\gamma$ can be estimated from
eq.~(\ref{eq:chi_def}) as
\begin{equation}
  \label{eq:l_gamma}
  l_\gamma = 2 \chiA \frac{\rhoC}{b \epsilon_\gamma}\,.
\end{equation}
Photon energy $\epsilon_\gamma$ depends on the particle energy
$\epsilon_\pm$ which depends on $\lPe$ according to
eq.~(\ref{eq:epsilonP_s}), and so $l_\gamma$ is a function of
$\lPe$. $\GAP{l}$ can be found by minimizing $l=\lPe+l_\gamma$ using
$\lPe$ as an independent variable: $\GAP{l}$ is the value of $l$ which
satisfies $dl/d\lPe=0$, $\GAParg{l}{\gamma}$ and $\lPeGAP$ are the
values of $l_\gamma$ and $\lPe$ where $l$ reaches its minimal value.
Using eq.~(\ref{eq:l_gap_def}) we can write an equation for
$\GAParg{l}{\gamma}$
\begin{equation}
  \label{eq:l_gamma_gap_eq}
  \frac{d\GAParg{l}{\gamma}}{d\lPeGAP}=-1 \,.
\end{equation}

If the photon energy depends on $\lPe$ as
$\epsilon_\gamma\propto{}(\lPe)^\alpha$, then
eq.~(\ref{eq:l_gamma_gap_eq}) is reduced to
\begin{equation}
  \label{eq:s_e_eq}
  \lPeGAP=\alpha{}\GAParg{l}{\gamma}\,,
\end{equation}
where $\GAParg{l}{\gamma}$ is expressed in terms of $\lPeGAP$
using eq.~(\ref{eq:l_gamma}). $\GAP{l}$ is then given by
\begin{equation}
  \label{eq:l_gap}
  \GAP{l} = \frac{\alpha+1}{\alpha} \lPeGAP\,.
\end{equation}
The final energy of the primary particle is given by
eq.~(\ref{eq:epsilonP_s}) with $\lP=\GAP{l}$.  Please note that
because the gap moves, the actual size of the gap (see
Fig.~\ref{fig:accelerator}(a)) is
\begin{equation}
  \label{eq:h_gap}
  \GAP{h}=\left(1+\frac{v}{c}\right)\GAP{l}\,.
\end{equation}

The energy of the CR photons depends on the particle energy as
$\epsilonP^3$ (eq.~\ref{eq:CR_ephot}), the particle energy depends on
$\lP$ as $\lP^2$ (eq.~\ref{eq:epsilonP_s}), hence,
$\epsilonG\propto{}(\lPe)^6$ and $\alpha=6$.  Substituting expression
for $\epsilonP$ (eq.~\ref{eq:epsilonP_s}) into the expression for CR
photon energy $\epsilon_\gamma$ (eq.~\ref{eq:CR_ephot}), the latter
into eq.~(\ref{eq:l_gamma}), and the resulting expression for
$l_\gamma$ into eq.~(\ref{eq:s_e_eq}) with $\alpha=6$ after algebraic
transformations we get the following expression for $\lPeGAP$
\begin{equation}
  \label{eq:s_e_free}
  \lPeGAP=\left(\frac{B_q^4 \lambdaC^2 c^3}{\pi^3} \right)^{1/7}
  \chiA^{1/7} \xi_j^{-3/7} \rhoC^{2/7}  P^{3/7} B^{-4/7}\,.
\end{equation}
The size of the gap as seen by the moving particle according to
eq.~(\ref{eq:l_gap}) is
\begin{equation}
  \label{eq:l_gap_free}
  \GAP{l} \simeq \sci{2}{4}\: \chiA^{1/7} \xi_j^{-3/7}
  \SUB{\rho}{c,\,7}^{2/7}\,
  P^{3/7} B_{12}^{-4/7} \mbox{cm} \,,
\end{equation}
where $B_{12}\equiv{}B/10^{12}\mbox{G}$ and
$\SUB{\rho}{c,\,7}\equiv{}\rhoC/10^7\mbox{cm}$.  Substituting
$\lP=\GAP{l}$ into eq.~(\ref{eq:epsilonP_s}) we get for the final
energy of particles accelerated in the gap%
\footnote{Our expression for the energy of primary particles has the
  same dependence on $\rhoC$, $P$ and $B$ as the expression for the
  potential drop in the gap derived by \citet{RudermanSutherland1975},
  their eq.~(23).  This is to be expected as in both cases particles
  are accelerated by the electric field which grows linearly with the
  distance and the size of the gap is regulated by absorption on
  curvature photons in magnetic field.  The difference is in the
  presence of factor $\xi_j$ and a different numerical factor.}
\begin{eqnarray}
  \label{eq:epsilonP_final}
  \SUBB{\epsilon}{\pm}{acc} & = & \frac{49}{18} \left( \frac{\pi B_q}{\lambdaC^3 c} \right)^{1/7}
                                   \chiA^{2/7}\; \xi_j^{1/7} \rhoC^{4/7} P^{-1/7} B^{-1/7} \nonumber \\
                       & \simeq & \sci{5}{7}\: \chiA^{2/7}\; \xi_j^{1/7} \SUB{\rho}{c,\,7}^{4/7} P^{-1/7} B_{12}^{-1/7}\,.
\end{eqnarray}

\begin{figure}
  \includegraphics[clip,width=\columnwidth]{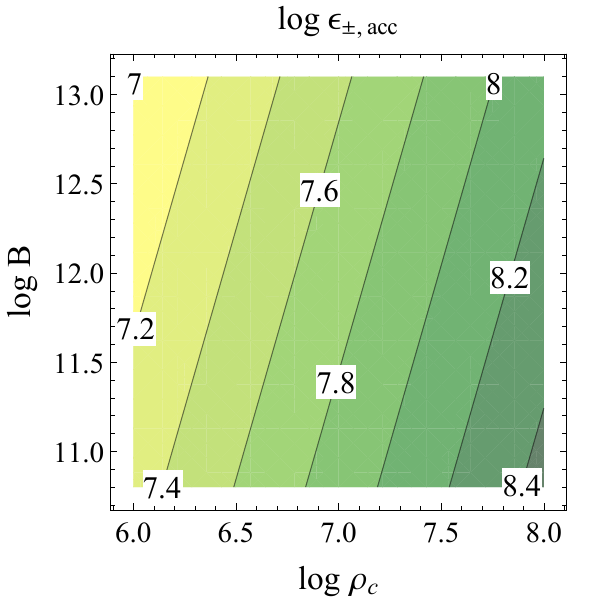} 
  \caption{Primary particle energy: contours of
    $\log\SUBB{\epsilon}{\pm}{acc}$ as a function of logarithms of the
    radius of curvature of magnetic field lines $\rhoC$ in cm and
    magnetic field strength $B$ in Gauss. We used the following values
    for gap parameters $P=33$~ms, $\xi_j=2$ and $\chiA=1/7$.}
  \label{fig:e_acc}
\end{figure}

The dependence of the primary particles energy on pulsar period $P$,
filling factor $\xi_j$ and the strength of the magnetic field $B$ is
very weak, the only substantial dependence is on the radius of
curvature of magnetic field lines.  The reason for this is the strong
dependence of CR photon energy on the energy of emitting particles
$\epsilonG\propto\epsilonP^3$, eq.~(\ref{eq:CR_ephot}).  Changes in
the threshold energy of pair producing photons which can stop the gap
growth cause only modest variation of the energy of primary particles,
which explicitly depends on $P$, $\xi_j$ and $B$, but not on $\rhoC$,
eq.~(\ref{eq:epsilonP_s}).  The energy of pair producing photons sets
the energy of accelerated particles, and in pulsars with a strong
accelerating electric field the gap will be smaller than in pulsars
with a weaker accelerating field.  In Fig.~\ref{fig:e_acc} we plot the
energy of particles $\SUBB{\epsilon}{\pm}{acc}$ accelerated in the gap
of a pulsar with $P=33$~ms as a function of the radius of curvature of
magnetic field lines $\rhoC$ and magnetic field strength $B$, assuming
$\xi_j=2$ and $\chiA=1/7$. The value $\chiA\approx1/7$ corresponds to
$\chiA$ of CR photons emitted by relativistic particles with
$\epsilonP^0=\sci{2.5}{7}$ in a magnetic field $B=10^{12}$~G with
$\rhoC=10^7$~cm, this is a good estimate for $\chiA$ in
eq.~(\ref{eq:epsilonP_final}) for young pulsars as the dependence on
$\chiA$ is very weak.  This plot clearly illustrates the dependence of
$\SUBB{\epsilon}{\pm}{acc}$ on $B$ and $\rhoC$ -- weaker magnetic
field and/or larger radius of curvature requires larger photon
energies for terminating gap growth, and so the energy of the primary
particles is larger.

\begin{figure}
  \includegraphics[clip,width=\columnwidth]{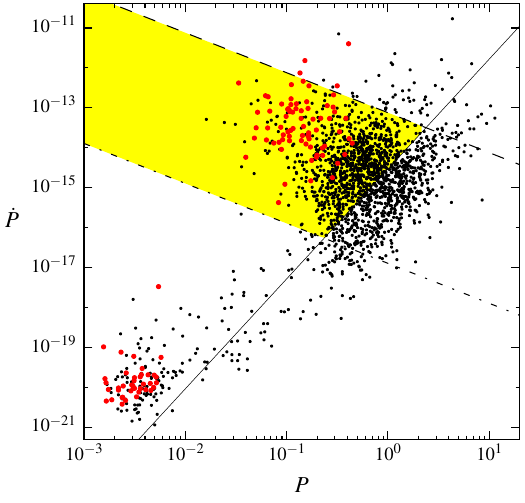}
  \caption{$P\dot{P}$ diagram with the yellow area showing the range
    of parameters where approximation for particle acceleration used
    in this paper is applicable, see text for description.  Pulsars
    from ATNF catalog (\citet{ATNF_Catalog2005},
    http://www.atnf.csiro.au/research/pulsar/psrcat) are shown by
    black dots, $\gamma$-ray pulsars from the second Fermi catalog
    \citep{Abdo2013_SecondPSRCatalog} by red dots.}
  \label{fig:ppdot}
\end{figure}

We derived eq.~(\ref{eq:epsilonP_final}) under the following
assumptions: (i) particles are accelerated freely, i.e. radiation
reaction can be neglected, (ii) the length of the gap is much smaller
that the polar cap radius, so that a one dimensional approximation can
be used, (iii) the magnetic field is
$B\lesssim{}0.2B_q\approx\sci{8.8}{12}$~G so that the opacity to
$\gamma{}B$ pair creation is described by eq.~(\ref{eq:alphaB}).
Constraints on the pulsar parameters (i) and (ii) are derived in
Appendix~\ref{sec:trans-radi-react} and Appendix~\ref{sec:limit-1-d}
correspondingly.  Plotted on the $P\dot{P}$ diagram,
Fig.~\ref{fig:ppdot}, these restrictions select the range of pulsar
period and period derivatives -- shown as yellow region -- where all
three assumptions are valid.  In this figure, the one-dimensional
approximation (ii) is valid to the left of the solid line, given by
eq.~(\ref{eq:1d-limit-dipole-PC_edge}), the approximation (i) of free
acceleration above the dot-dashed line, given by eq.~(\ref{eq:B_rr}),
and pulsars with $B<0.2B_q$ are below the dotted line.  We see that
most of young normal pulsars, including gamma-ray pulsars from the
Fermi second pulsar catalog, fall in this range.

\section{Cascade multiplicity per primary particle: 
  semi-analytical model}
\label{sec:mult-semi-analytical}

\begin{figure*}
  \label{fig:multiplicity-with-acc}
  \includegraphics[clip,width=\textwidth]{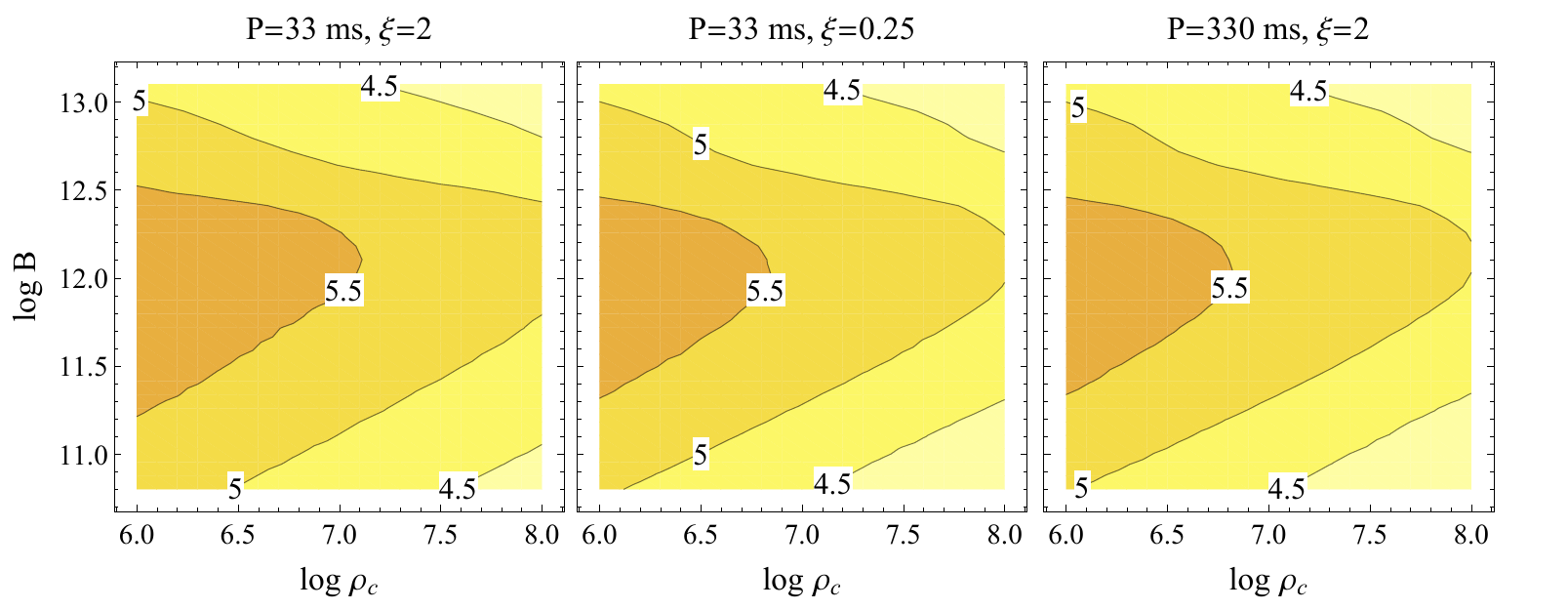}  
  \caption{Multiplicity of polar cap cascades: contours of
    $\log\kappa$ as a function of logarithms of curvature of magnetic
    field lines $\rhoC$ in cm and magnetic field strength $B$ in Gauss
    for three sets of the gap parameters $(P[\mbox{ms}],\xi_j)$:
    $(33,2)$, $(33,0.25)$, $(330,2)$. In all cases we used $\chiA=1/7$
    for calculation of the energy of primary particles.}
\end{figure*}

We now combine the results of \S\ref{sec:mult-CR-synch} concerning the
cascade multiplicity for a fixed energy of the primary particle, and
of \S\ref{sec:energy-primary} concerning the energy of primary
particles accelerated in the gap with given parameters $P$, $\xi_j$,
strength $B$, and $\rhoC$.  The multiplicity of CR-synchrotron
cascades depends on the energy of the primary particle $\epsilonP$,
magnetic field $B$, and radius of curvature of magnetic field lines
$\rhoC$.  The only significant dependence of the energy of accelerated
particles $\epsilonP$ in young pulsars is on the radius of curvature
of magnetic field lines $\rhoC$; the dependence on $P$, $\xi_j$, and
$B$ is very weak, see eq.~(\ref{eq:epsilonP_final}).  Therefore, when
particle acceleration is taken into account, the overall cascade
multiplicity $\kappa$ can depend substantially only on $B$ and
$\rhoC$.

In CR-synchrotron cascade changes in $\epsilonP$ and $\rhoC$ change
$\kappa$ in opposite directions -- for higher $\epsilonP$ multiplicity
is higher, for larger $\rhoC$ multiplicity is lower.  But the energy
of the primary particles accelerated in polar caps of young pulsars
$\epsilonP$ is higher for larger radii of curvature of magnetic field
lines $\rhoC$.  Hence, increasing $\rhoC$ lowers the cascade
multiplicity for a fixed $\epsilonP$, but at the same time increases
$\epsilonP$, which partially compensates the decrease of cascade
multiplicity.  Therefore, the final cascade multiplicity should have a
rather weak dependence on $\rhoC$, which leaves the magnetic field
strength $B$ the only parameter significantly affecting the
multiplicity of strong cascades in polar caps of young pulsars.

In Fig.~\ref{fig:multiplicity-with-acc} we show the final multiplicity
as a function of magnetic field strength and radius of curvature of
magnetic field lines.  We present plots for three sets of parameters
$P$ and $\xi_j$ which differ by $\sim$ an order of magnitude.  As
expected, both pulsar period as well as the filling parameter $\xi_j$
have a very small effect on the final multiplicity.  For the range of
$\rhoC$ and $B$ in Fig.~\ref{fig:multiplicity-with-acc}, the
CR-synchrotron cascade has the highest efficiency -- primary particles
lose most of their energy as CR photons within the distance
$\CR{s}=\RNS$ where the cascade is possible -- and higher
multiplicities cannot be achieved.  
According to Fig.~\ref{fig:multiplicity-with-acc} the cascade multiplicity
scales with $\rhoC$ roughly as $\kappa\propto(\rhoC)^\nu$, with
$\nu\lesssim1/2$. The interval $[10^6,10^8]$cm represents the most reasonable
values of $\rhoC$ for any global magnetic field configuration in the pulsar
polar cap. For very different magnetic field configurations -- a highly
multipolar field with $\rhoC\sim10^6$cm vs. a dipole field with
$\rhoC\sim10^8$cm -- the multiplicity differs by less than an order of
magnitude. The dependence of $\kappa$ on the magnetic field is stronger, with
the maximum reached near $B\sim10^{12}$~G.

It is a remarkable fact that the multiplicity of the most efficient cascades is
sensitive mostly to the strength of the magnetic field.  The multiplicity is not
very sensitive to $\rhoC$ and for typical pulsar magnetic field of $10^{12}$~G
is around $10^5$.  For fixed $B$ the total pair yield -- the total number of
particles injected into the magnetosphere -- depends then only on the flux of
primary particles.  This is true for pulsars where the accelerating potential is
regulated by pair production and where cascade operate in CR-synchrotron regime
have high efficiency.

\section{Cascade multiplicity per primary particle: 
  Numerical simulations}
\label{sec:mult-numerical}

In \Ss\ref{sec:photon-absorption}--\ref{sec:mult-semi-analytical} we
developed a semi-analytical model of strong polar cap cascades. In
order to verify the key assumptions and conclusions of this model we
have performed numerical simulations of time-dependent polar cap pair
cascades. Because such numerical simulations are quite time consuming
we limited ourselves to the case of a young pulsar with parameters
similar to the Crab pulsar ($P=33$~ms) and magnetic field strength $B$
and radius of curvature of magnetic field lines $\rhoC$ having values
resulting in high multiplicity.  Our goal was to show that the main
assumptions and conclusions of our analysis of polar cap cascades are
realistic.  More extensive self-consistent numerical studies of the
polar cap cascades will be done in subsequent papers.

As outlined in \S\ref{sec:overview-particle-acceleration}, particles
are quickly accelerated in the gap which is much smaller than the
typical distance over which the full cascade develops.  The primary
pair producing particles are moving most of the time in the region
with screened electric field. \emph{If} the primary particle energies
are known, the full cascade can be modeled using traditional
Monte-Carlo techniques \citep{Daugherty/Harding82}; to obtain the
energies of primary particles initiating the CR-synchrotron cascade a
self-consistent model of the cascade
\citep{Timokhin2010::TDC_MNRAS_I,TimokhinArons2013} is necessary.
  
We have performed numerical simulations of time-dependent polar cap
pair cascades in a two-step process.  In the first step, we use a
hybrid Particle-in-Cell/Monte Carlo (PIC/MC) code PAMINA (\textbf{P}IC
\textbf{A}nd \textbf{M}onte-Carlo code for cascades \textbf{IN}
\textbf{A}strophysics) to simulate the initial self-consistent
electric field generation, particle acceleration and electric field
screening near the NS to obtain the distribution functions of the
electrons and positrons in the acceleration/screening region.  This
code includes only CR of the particles and first generation of pairs
needed to screen the gap, and so does not follow the full synchrotron
cascade.  The details of this code are described in
\citet{Timokhin2010::TDC_MNRAS_I,TimokhinArons2013}.

In the second step, we use another code to simulate the full pair
cascade in the pulsar dipole field above the PC, including both CR of
primary particles and synchrotron radiation of pairs.  This code,
based on the calculation described in detail in
\citet{HardingMuslimov2011} [HM11], is a Monte-Carlo simulation of the
electron-positron pair cascade generated above a PC by accelerated
particles in the region of screened electric field.  Although HM11
included a steady particle acceleration component, this component is
not used in the present calculation.  We therefore assume that the
particles and the further pairs they create do not undergo any
acceleration.  This code takes as input the distribution functions of
accelerated particles output by the time-dependent PAMINA code that
are moving away from the NS surface and simulates the combined cascade
from all of the particles.  Although our setup is capable of
calculating full cascades generated by primary particles with
arbitrary distribution functions, in the simulations described in this
section we used a monochromatic injection of primary particles. On the
one hand, the energy distribution of the most energetic primary
particles which produce the bulk of the pairs in many cases is close
to monochromatic (see e.g.  \S\ref{sec:flux-primary},
Fig.~\ref{fig:xp_xpdensity_superGJ}), and on the other hand this
enables us to compare numerical simulations with predictions of our
semi-analytical theory.  The energy of the primaries was calculated
from the self-consistent model however.

The MC code first follows the primary particle in discrete steps along
the magnetic field line at magnetic colatitude $\theta$, starting from
the location $x_0$ and particle energy $\epsilonP^0$ at time
$\SUB{t}{peak}$ at the peak of the pair production cycle in PAMINA
code, computing its curvature radiation.  The steps $\Delta{}x$ are
set to the minimum of a fraction $0.1$ of a NS radius and the distance
over which the particle would lose 1\% of its energy to curvature
radiation.  Dividing the CR spectrum at each step into logarithmic
energy intervals, a representative photon from each energy interval is
followed through the curved magnetic field until its point of pair
production (determined as a random fraction of the mean-free path).
The number of CR photons in each energy bin, $\CR{n}$, is determined
by the energy loss rate and average energy in that bin.  The pairs
produced by the photon, or the escaping photon number, is then
weighted by $\CR{n}$.  The created pair is assumed to have the same
direction and half the energy of the parent photon.  Although the CR
photons are radiated parallel to the magnetic field, they must acquire
a finite angle to the field before producing a pair, so the created
pairs have finite pitch angles at birth.  Each member of the pair
emits a sequence of cyclotron and/or synchrotron photons, starting
from its initial Landau state until it reaches the ground state,
assuming the position of the particle remains fixed (given the very
rapid radiation rate).  As described in HM11, when the pair Landau
state is larger than 20, the asymptotic form of the quantum
synchrotron rate \citep{SokolovTernovBook68} is used to determine the
photon emission energy and final Landau state.  When the Landau state
is below 20, the full QED cyclotron transition rate
\citep{HardingPreece1987} is used.  At large distances above the NS
surface, when the magnetic field drops below $0.002 B_q$, we short-cut
the individual emission sequence and use an expression for the
spectrum of synchrotron emission for an electron that loses all of its
perpendicular energy \citep{Tademaru1973}.  Each emitted photon is
then propagated through the magnetic field from its emission point
until it pair produces or escapes.  The next generation of pairs are
then followed through their synchrotron/cyclotron emission sequence.
By use of a recursive routine that is called upon the emission of each
photon, we can follow an arbitrary number of pair generations.  The
cascade continues until all photons from each branch have escaped.  As
each member of each created pair completes its synchrotron emission,
its ground-state energy, position and generation number are stored in
a pair table.  As each photon either pair produces or escapes, its
energy, generation and position of pair creation or escape are stored
in tables for absorbed and escaping photons.  The photons and pairs
from all accelerated particles are summed together to produce the
complete cascade portrait at that time step.

The NS magnetic field in the MC code described above is a distorted
dipole with an azimuthal ($\phi$) component which is off-set from the
center of the NS.  The magnetic field is given by
\begin{eqnarray} 
  \mathbf{B} & = & B_0\left(\frac{\RNS}{r}\right)^3\times \nonumber \\
             &   & \left[ \mathbf{e_r}\cos[\theta(1+a)] +
                \mathbf{e_\theta}\;\frac{1}{2}\,\sin[\theta(1+a)]
                \right. \nonumber                                   \\
             &   & \phantom{[}\left. -\mathbf{e_\phi}\;\frac{1}{2}\,\varepsilon (\theta
                + \sin\theta\cos\theta )\,\sin(\phi - \phi_0) \right], 
\label{eq:B2}
\end{eqnarray}
where $B_0$ is the surface magnetic field strength at the magnetic
pole, $r$ is the radial coordinate,
$a=\varepsilon\;\cos(\phi - \phi _0)$ is the parameter characterizing
the distortion of polar field lines, and $\phi_0$ is the magnetic
azimuthal angle defining the meridional plane of the offset PC.  The
parameter $\varepsilon$ sets the magnitude and the parameter $\phi_0$
sets the direction of asymmetry of this azimuthal component.  Setting
$\varepsilon = 0$ gives a pure dipole field structure, while a
non-zero value of $\varepsilon$ produces an effective offset of the PC
from the dipole axis in the direction specified by $\phi_0$.  For
non-zero $\varepsilon$, the radius of curvature of the magnetic field
lines is smaller than dipole in the direction of the offset and larger
than dipole in the direction opposite to the direction of offset.
This particular parametric form for the magnetic field was used in
simulations of stationary cascades in HM11, it was chosen to account
for distortion of the shape of the PC caused by currents flowing in
pulsar magnetosphere.  Such azimuthal asymmetries in the near-surface
magnetic field is caused by the sweepback of the field lines near the
light cylinder due to retardation \citep[e.g.][]{Dyks/Harding04} and
currents
\citep[e.g.][]{Timokhin2006:MNRAS1,BaiSpitkovsky2010a,Kalapotharakos2014}
or, additionally, by asymmetric currents in the NS.

We did not perform a systematic study of all parameter space with our
numerical simulations, which will be done elsewhere, with our
numerical simulations we test assumptions and predictions of our
semi-analytic cascade model.  Any form of magnetic field with
adjustable radius of curvature of magnetic field lines would serve our
purposes, but using the magnetic field given by eq.~(\ref{eq:B2})
allows comparison with the most recent simulations in the frame of the
previous-generation cascade models HM11.  We explored cases of pair
cascades both for pure dipole fields and for azimuthally distorted
fields.  Multiplicities obtained from the numerical simulations agree
reasonably well with the semi-analytic model, within a factor of a
few.  As an example we describe in detail results of particular
simulations with pulsar parameters yielding high multiplicity for a
cascade at the peak of the pair creation cycle. The magnetic field is
$B=10^{12}$~G and is moderately distorted, with the offset
$\varepsilon=0.4$ resulting in the radius of curvature of magnetic
field lines near the NS $\rhoC=\sci{8.8}{6}$~cm. The initial energy of
primary particles from PAMINA simulations is
$\epsilonP^0=\sci{2.3}{7}$.

\begin{figure*}
  \includegraphics[clip,width=\textwidth]{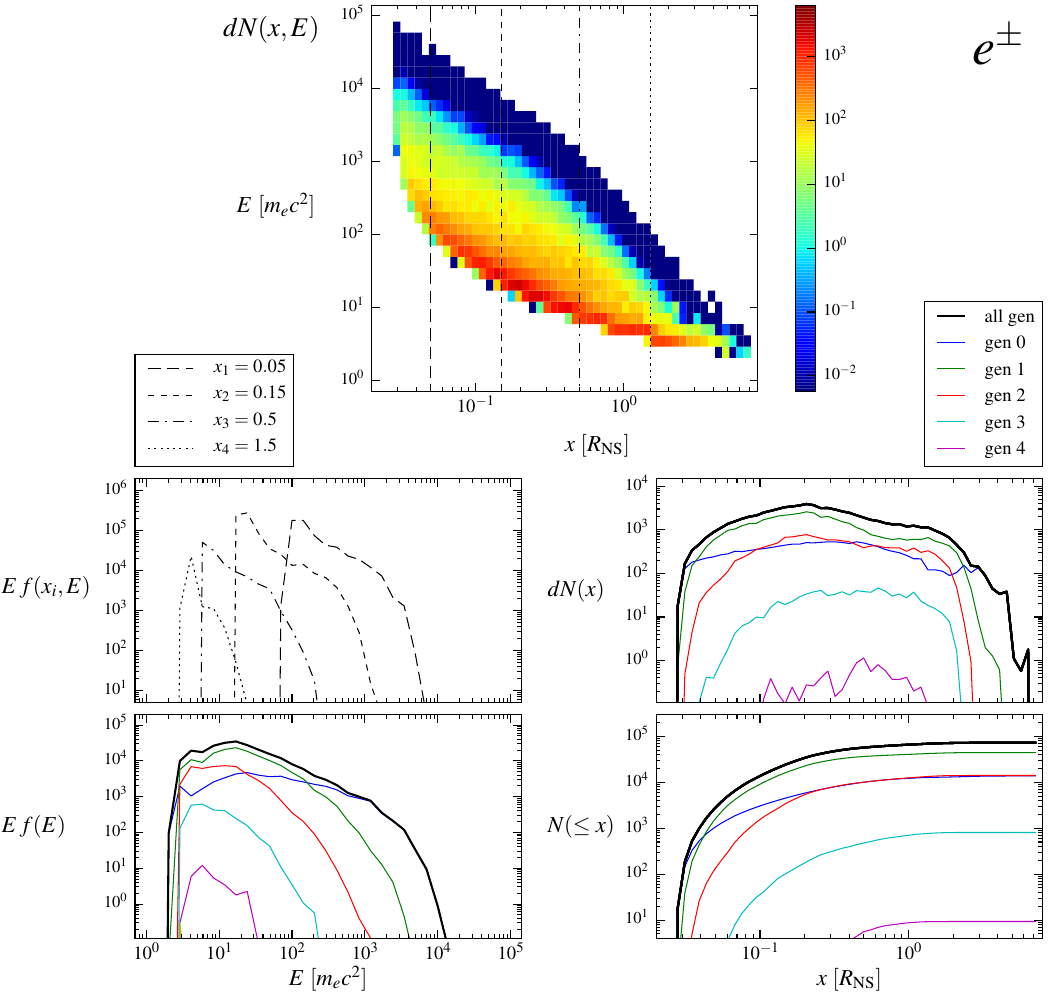}
  \caption{Cascade portrait of electron-positron pairs.
    \textit{Top panel}:
    number of particles produced in each energy and distance bin $dN(x,E)$ color coded in
    logarithmic scale according to the colorbar on the right.  
    \textit{Middle left}: $E\,f(x_i,E)$ -- energy distribution  of particles produced at four
    distances $x_i,i=1\dots4$; different line styles correspond to
    different distances according to the left plot legend. 
    \textit{Bottom left}: $E\,f(E)$ -- energy distribution of all particles
    produced in the cascade.
    \textit{Middle right}: $dN(x)$ -- differential pair production rate
    -- number of particles produced in a distance bin.
    \textit{Bottom right}: $N(\le{}x)$ -- total number of
    particles produced up to the distance $x$.
    Color lines in plots for $E\,f(E)$, $dN(x)$, and $N(\le{}x)$ show contributions of
    different cascade generations, lines are color-coded according to the
    right plot legend.  Thick black lines show contributions of all
    cascade generations.   
    $x$ is the distance from the NS normalized to NS radius $\RNS$ and
    $E$ is particle energy normalized to $m_ec^2$.  Particle number
    density is normalized to $\GJ{n}$.  Parameters of this simulation:
    pulsar period $P=33$~ms; the magnetic field in the PC has
    $B_0=10^{12}$~G and $\rhoC=\sci{8.8}{6}$~cm; initial energy of
    primary particles $\epsilonP^0=\sci{2.3}{7}$. }
  \label{fig:portrait-pairs}
\end{figure*}

\begin{figure*}
  \includegraphics[clip,width=\textwidth]{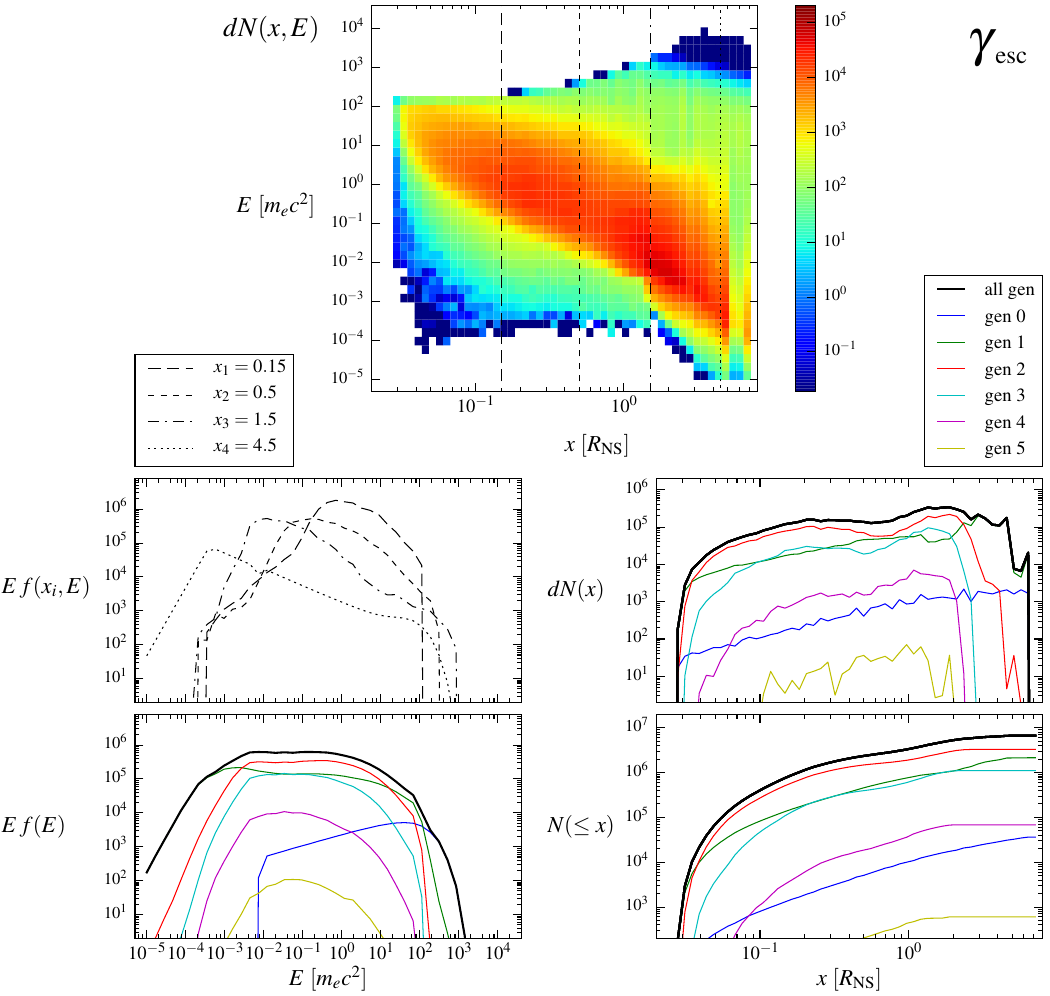}
  \caption{Cascade portrait of escaping photons for the same cascade
    as in Fig.~\ref{fig:portrait-pairs}. Notations and normalizations
    are the same as in Fig.~\ref{fig:portrait-pairs}.}
  \label{fig:portrait-photons}
\end{figure*}

In Figures~\ref{fig:portrait-pairs} and \ref{fig:portrait-photons} we
show a set of five plots which we call ``cascade portraits''.  These
plots illustrate different aspects of the cascade development by
showing moments of the photon or particle distribution function
$f(x,E)$.  The top panel shows the number of particles produced in
each energy and distance bin
\begin{equation}
  dN(x,E)=f(x,E)\,dx\,dE,
\end{equation}
as a 2D color map.  The number of particles is color coded in
logarithmic scale according to the color bar on the right.  The middle
left plot shows the energy distribution $E\,f(x_i,E)$ of particles
produced at four distances $x_i,i=1\dots4$; different line styles
correspond to different distances according to the left plot legend.
These spectra are essentially cross-sections of the map of $dN(x,E)$
(multiplied by particle energy) along four lines shown in the plot for
$dN(x,E)$. The bottom left plot shows the energy distribution of all
particles produced in the cascade ($dN(x,E)$ integrated along $x$
direction and multiplied by $E$)
\begin{equation}
  E\,f(E)=E\int_0^{x_{\max}}d\tilde{x}\,f(\tilde{x},E)\,.
  \label{eq:EfE}
\end{equation}
In this and the following plots, colored lines show contributions of
different cascade generations; lines are color-coded according to the
right plot legend.  The middle right plot shows the differential pair
production rate -- number of particles produced in a distance bin
($dN(x,E)$ integrated along the $E$ direction)
\begin{equation}
  dN(x)=dx\,\int_0^{E_{\max}}f(x,E)\,dE\,.
  \label{eq:dNx}
\end{equation}
The bottom right plot shows the cumulative pair production rate --
total number of particles produced up to the distance $x$
\begin{equation}
  N(\le{}x)=\int_0^xd\tilde{x}\,\int_0^{E_{\max}}f(\tilde{x},E)\,dE\,.
  \label{eq:Nx}
\end{equation}

Figure~\ref{fig:portrait-pairs} shows the cascade portrait of the
pairs (we do not differentiate between electrons and positrons).  The
cascade extends to about $\sim6\RNS$, where it dies out completely as
the magnetic field strength and the accelerated particle energy
decrease.  The pair number grows very quickly, within a few tenths of
a NS radius, and then saturates at $\kappa\approx\sci{7.4}{4}$.  The
majority of pairs are produced at distances $<\RNS$ (see plot for
$N(\le{}x)$), which supports the assumption about the length of the
cascade zone being $\sim\RNS$ made in \S~\ref{sec:opacity-gamma-b}.
The number of pair generations is highest at $\lesssim\RNS$, up to six
in this case%
\footnote{There are too few pairs produced in the 6th generation to
  show in the plot; curves corresponding to this generation are below
  the lower limit of all plots in Fig.~\ref{fig:portrait-pairs}; this
  generation shows in the portrait for photons,
  Fig.~\ref{fig:portrait-photons}}.
The occurrence of most of the cascade generations at distances up to
$\sim\RNS$ is the evidence that the photon mfp in a strong cascade is
indeed small, $l_\gamma\ll\RNS$, so that the cascade initiated by any
given particle goes through several generations within the distance
$\sim\RNS$.  The large extent of the cascade zone relative to
$l_\gamma$ is due to continuous injection of pair-producing CR photons
(see the blue line in the plot of $dN(x)$; generation 0 pairs are
produced by CR photons).  The largest contribution to pair
multiplicity in this case comes from generation 1, pairs created by
the first synchrotron photons (see plots for $dN(x)$ and $N(\le{}x)$).
In our simulations for different pulsar parameters, contributions of
generation 1 and 2 to the pair multiplicity sometimes become
comparable; for weaker cascades the relative contribution of
generation 0 is higher than in this case, however we did not see
generations 3 and higher producing the majority of pairs.  The number
of cascade generations is not very large in any of our simulations
(several at most), but in each generations high numbers of pair
producing photons are emitted which results in high multiplicity.

The energy of created pairs decrease with distance (see plots for
$dN(x,E)$ and $E\,f(x_i,E)$), mostly because of energy losses of
primary particles which results in lower energy CR photons.  The pair
spectrum extends down to a few $mc^2$, since the cascade is very
efficient at converting initial pair energy into more photons (and
pairs).  Degradation of pair energies through cascade generations
discussed in \S~\ref{sec:mult-synchr-casc} is clearly visible on the
plot for $E\,f(E)$ -- the maximum pair energy systematically decreases
with cascade generations at all distances.

Figure ~\ref{fig:portrait-photons} show the portrait of photons
escaping the cascade.  Photon generation 0 is CR while the higher
generations ($\le 1$) are synchrotron/cyclotron radiation.  Although
the highest energy photons are produced nearest the NS surface, these
photons are absorbed by pair production attenuation so that the
spectra at the lowest altitudes show sharp cutoffs near
$100~mc^2$. This cutoff is clearly visible in the plot for $Ef(x_i,E)$
for $x_1=0.15\RNS$ shown by a dashed line.  In the plot for $dN(x,E)$
the cutoff is evident as a sharp horizontal boundary of the colored
region for $x\lesssim0.16\RNS$.  The escaping photon energies increase
with distance from the NS surface, as the magnetosphere becomes more
transparent.  The highest escaping photon energies are produced near
the end of the cascade, at around 4 NS radii. The highest energy
photons escaping the cascade are CR photons (blue line in the plot for
$Ef(E)$).  Synchrotron radiation is emitted by pairs right after their
creation, so that the pair formation and synchrotron radiation end at
the same distance -- in the plot for $dN(x)$, the number of photons in
each generation drops at large distances in accordance with the drop
of number of injected pairs shown on a similar plot in
Fig.~\ref{fig:portrait-pairs}.  Above $\sim4\RNS$ the emitted CR
photon energies drop as the primary particles continue to lose energy.
The spectrum of escaping photons also broadens at the lower end
because, as the magnetic field decreases, so does the cyclotron energy
which sets the lower limit of the synchrotron spectrum.  Thus the
lowest and the highest energy escaping photons are produced at the
largest distances from then NS, cf. spectra at different $x_i$ in the
plot for $Ef(x_i,E)$. The bulk of high energy emission from the PC
cascade comes from synchrotron radiation of pairs in generations 1,2,
and 3.

Quantitatively, our semi-analytic theory compares with this particular numerical
simulation as follows.  For pulsar parameters used in this simulation, the
energy of primary particles according to eq.~(\ref{eq:epsilonP_final}) should be
$\epsilonP^{0,\,\rmscriptsize{a}}\approx\sci{3.6}{7}$, which is $\approx1.6$
times larger than the result obtained from numerical simulations with PAMINA
code $\epsilonP^{0,\,\rmscriptsize{num}}\approx\sci{2.3}{7}$.  For the
multiplicity of the CR-synchrotron cascade started by primary particles with
monochromatic energies $\epsilonP^{0,\,\rmscriptsize{num}}$, the semi-analytic
model gives $\CRsyn{\kappa}^{\rmscriptsize{a}}\approx\sci{1.56}{5}$, from
eq.~(\ref{eq:kappa-CR-Sync}), which is also about $2$ times higher than the
value obtained in numerical simulations
$\kappa^{\rmscriptsize{num}}\approx\sci{7.4}{4}$.  The combined model from
\S~\ref{sec:mult-semi-analytical}, which uses the analytic model for particle
acceleration as an input for the semi-analytic model of CR-synchrotron cascade,
predicts for the multiplicity $\kappa^{\rmscriptsize{a}}\approx\sci{2.9}{5}$,
see Fig.~\ref{fig:multiplicity-with-acc}, which is $\approx4$ times larger than
the multiplicity from numerical simulations.  The discrepancy with the
semi-analytic model for several other numerical simulations we performed with
different parameters is of the same order. We attribute this discrepancy mostly
to the approximation of constant magnetic field -- in the numerical simulations,
where $B$ and $\rhoC$ (which is derived from $B$) depend on the distance
according to eq.~(\ref{eq:B2}), photon absorption decreases and becomes less
efficient with the distance. We think that for such a simple model the agreement
with the numerical simulations is reasonable and the model can be used for
estimates of multiplicities in young energetic pulsars.

\section{Resonant Inverse Compton Scattering}
\label{sec:reson-inverse-compt}

\begin{figure}[t]
  \centering
  \includegraphics[clip,width=\columnwidth]{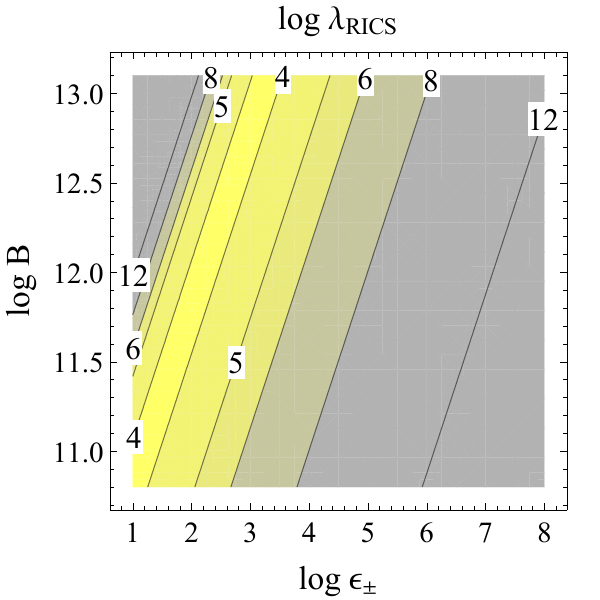}
  \caption{Distance (in cm) over which a particle loses its energy via
    resonant inverse Compton scattering. Contours of
    $\log{\SUB{\lambda}{RICS}}$ are plotted as a function of
    logarithms of particle energy $\epsilonP$ and magnetic field $B$
    in Gauss for a NS surface temperature $T=10^6$K and $\mu_s=0.5$.}
  \label{fig:Lrics}
\end{figure}

\begin{figure}[t]
  \centering
  \includegraphics[clip,width=\columnwidth]{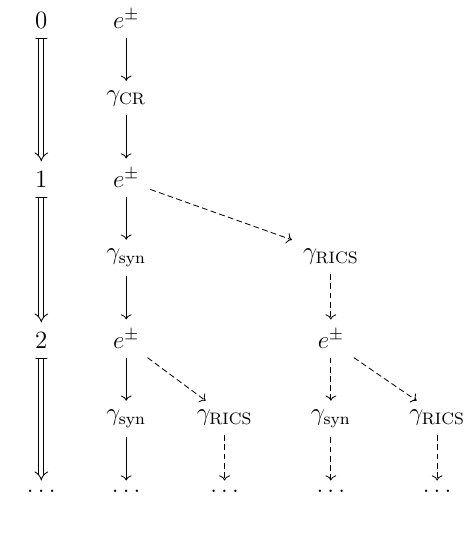}
  \caption{Diagram showing the general chain of physical processes in
    a strong polar cap cascade. Cascade generations are shown on the
    left -- numbers connected by double arrows.  In each generation
    particles $e^\pm$ (electrons and/or positrons) produce photons
    ($\CR{\gamma}$ -- via curvature radiation, $\SYN{\gamma}$ -- via
    synchrotron radiation, $\SUB{\gamma}{RICS}$ -- via Resonant
    inverse Compton scattering), which are turned into pairs of the
    next cascade generation.  The CR-synchrotron cascades studied in
    detail in this paper are shown by solid arrow, dashed arrows show
    RICS initiated branches which are discussed only in
    \S~\ref{sec:reson-inverse-compt}.   }
  \label{fig:cascade-diagram}
\end{figure}

\begin{figure}[t]
  \centering
  \includegraphics[clip,width=\columnwidth]{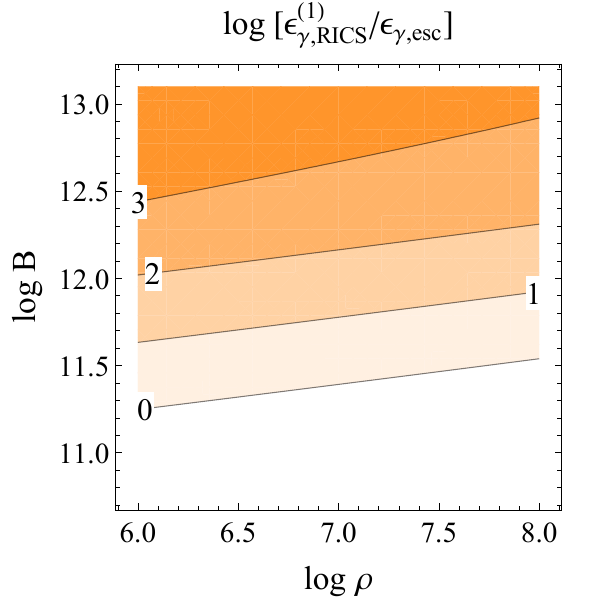}
  \caption{Ratio of of the characteristic energy of RICS photons
    $\epsilonGRICS^{(1)}$ emitted by the first
    generation pairs to the energy of escaping photons: contours of
    $\log\left[\epsilonGRICS^{(1)}/\epsilonGesc\right]$
    are plotted as a function of of logarithms of the radius of curvature
    of magnetic field lines $\rhoC$ in cm and magnetic field strength
    $B$ in Gauss. We used the following values for gap parameters
    $P=33$~ms, $\xi_j=2$ and $\chiA=1/7$. }
  \label{fig:fEricsEesc_1gen}
\end{figure}

Another emission mechanism for relativistic particles besides curvature and
synchrotron radiation is inverse Compton scattering (ICS).  In strong magnetic
fields typical for pulsar polar caps, ICS can occur in the resonant regime, when
the photon energy in the electron's rest frame is equal to the cyclotron energy.
The cross-section for scattering of such photons is greatly enhanced compared to
that of non-magnetic scattering.  It has been noted that resonant ICS (RICS)
with the soft thermal photons from the NS surface is important for high-energy
emission from pulsar polar caps \citep{Sturner1995,Zhang2000} and can be
important in the development of polar cap cascades because for quite a wide
range of pulsar parameters, scattered photons can be above pair-formation
threshold \citep[e.g.][]{SturnerDermerMichel1995,Zhang2000}.  In this section we
argue that although RICS can play a role in the development of polar cap
cascades, it never becomes the dominant source for pair multiplicity and,
therefore, considering only CR-synchrotron cascades provides adequate estimates
for pair multiplicity in strong cascades of normal pulsars. A detailed study of
the role of RICS in polar cap cascades will be presented in a subsequent paper.

First, let us consider the efficiency of RICS in transforming particle
kinetic energy into radiation.  The distance over which a particle
loses most of its energy to RICS is given by
\citep{Zhang2000,Sturner1995,Dermer1990}:
\begin{eqnarray}
  \SUB{\lambda}{RICS} & = & -0.061\, \epsilonP^2\, T_6^{-1} B_{12}^{-2}
                            \times \nonumber\\
                      && \ln^{-1}\left[1 - \exp\left( -\frac{134
                         B_{12}}{\epsilonP\, T_6 (1 -
                         \mu_s)}\right)\right] \, \mbox{cm}
  \label{eq:Lrics}
\end{eqnarray}
where $T_6$ is the temperature of the NS surface in units of $10^6$K, $B_{12}$
is the magnetic field strength in units of $10^{12}$~G, and
$\mu_s=\cos\theta_s$, where $\theta_s$ is the angle between the momenta of the
scattering photon and particle in the lab frame.  If the NS is young, its
surface temperature comes mostly from cooling and should be about $10^6$K.  It
emits X-ray photons necessary for RICS from the whole surface, and for particles
at a distance comparable to the NS radius the range of $\mu_s$ is quite large.
For older NS, with full surface temperatures below few$\times10^5$K only the
polar cap region, heated by the backflow of accelerated particles up to
$\sim\mbox{few}\times{}10^6$K
\citep[e.g.][]{Harding/Muslimov:heating_1::2001,Harding/Muslimov:heating_2::2002},
can emit enough photons for RICS to become important.  In the latter case when
the particle reaches a distance comparable to the width of the polar cap
$\PC{r}\simeq1.4\times10^4\,P$~cm, where $P$ is pulsar period in sec, the range
of $\mu_s$ gets very small and photons quickly get out of resonance.  So, for
young NSs RICS is an important radiation process if
$\SUB{\lambda}{RICS}\sim\RNS$; for old NSs this condition changes to
$\SUB{\lambda}{RICS}\sim\PC{r}$

In Fig.~\ref{fig:Lrics} we plot $\SUB{\lambda}{RICS}$ as a function of
particle energy $\epsilonP$ and magnetic field $B$, for $T=10^6$K and
$\mu_s=0.5$.  It is evident from this plot that primary particles,
with $\epsilonP>10^6$, lose a negligible amount of their energy via
RICS, and curvature radiation is the dominant emission mechanism for
generation 0 of strong cascades. For cold NSs, RICS of photons from
heated polar caps might be an important emission mechanism only for a
very narrow energy range of low energetic particles, in one of the
later cascade generation -- the parameter space between two
$\SUB{\lambda}{RICS}=10^4$ contours is rather small -- and for most
values of $B$ the scattered photons will be below pair formation
threshold.  Hence, RICS can be completely neglected in strong polar
cap cascades of cold NSs.  For hot NSs RICS becomes an important
emission mechanism for a wide range of moderate particle energies
$\epsilonP<10^4$ -- the energy range between contours of
$\SUB{\lambda}{RICS}=10^6$ in Fig.~\ref{fig:Lrics} is quite wide.  In
the latter case, the diagram for physical processes in a strong polar
cap cascade can have the general form shown in
Fig.~\ref{fig:cascade-diagram}, with RICS photons in some cases
carrying non-negligible energy starting from cascade generation 1.
The extension of the semi-analytical analysis of CR-synchrotron
cascade developed in
\Ss\ref{sec:photon-absorption}--\ref{sec:mult-semi-analytical} to the
whole cascade taking into account multiple cascade branches is not
straightforward and we postpone it to future publications.  Here we
will make rough estimates of the contribution of RICS cascade branches
to the multiplicity of the whole cascade.

Let us now discuss the efficiency of RICS cascade branches in splitting the
available energy into pairs.  The energy of photons after scattering by
particles of energy $\epsilonP$ in the RICS regime is \citep[e.g.][]{Zhang2000}
\begin{equation}
  \label{eq:e_RICS}
  \epsilonGRICS = 2 \SUBB{\epsilon}{\pm}{\textsc{f}}\, b\,,
\end{equation}
where $\SUBB{\epsilon}{\pm}{\textsc{f}}$, given by
eq.~(\ref{eq:gamma_F_chi}), is the kinetic energy of a particle moving
along a magnetic field line -- the final energy of freshly created
pairs after they emit all perpendicular to $B$ energy via synchrotron
radiation.  Using eqs.~(\ref{eq:e_RICS}),~(\ref{eq:gamma_F_chi})
together with eq.~(\ref{eq:CR_ephot}) for the energy of CR photons and
eq.~(\ref{eq:epsilonP_final}) for the energy of the primary particle,
we get an upper limit on the energy of the generation 1 RICS photons
-- the highest energy RICS photons.  In Fig.~\ref{fig:fEricsEesc_1gen}
we plot the ratio of the energy of generation 1 RICS photons
$\epsilonGRICS^{(1)}$ to the energy of photons escaping from the
cascade zone $\ESC{\epsilon}$ as a function of the radius of curvature
of magnetic field lines and the strength of the magnetic field $B$.
It is easy to see that for magnetic fields weaker than a few
$\times10^{11}$~G, even the highest energy RICS photons are not
capable of producing electron-positron pairs.  For stronger magnetic
fields, however, RICS photons do contribute to pair multiplicity.

\begin{figure}[t]
  \centering
  \includegraphics[clip,width=\columnwidth]{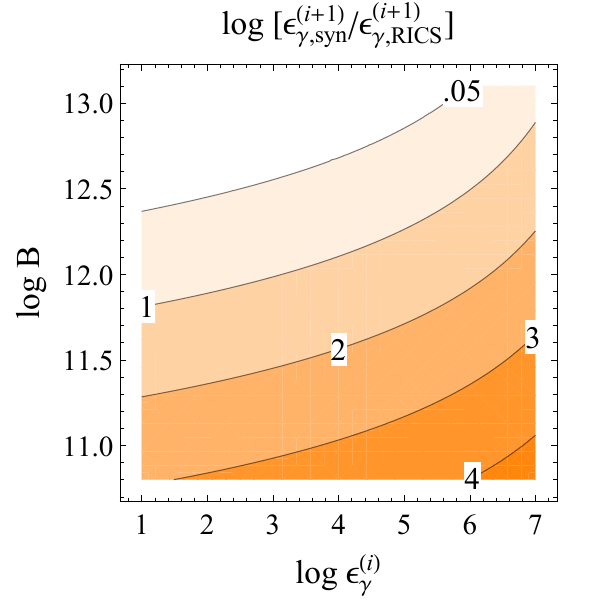}
  \caption{Ratio of the characteristic energies of synchrotron
    $\SUBB{\epsilon}{\gamma}{syn}^{(i+1)}$ and RICS
    $\epsilonGRICS^{(i+1)}$ photons emitted by freshly
    created pairs ($(i+1)^{th}$ generation cascade photons): contours
    of
    $\log\left[\SUBB{\epsilon}{\gamma}{syn}^{(i+1)}/\epsilonGRICS^{(i+1)}\right]$
    are plotted as a function of logarithms of the parent photon
    energy $\epsilonG^{(i)}$ and magnetic field strength $B$ in Gauss
    for $\rhoC=10^7\mbox{cm}$. }
  \label{fig:fEsynErics}
\end{figure}

The characteristic energy of RICS photons in terms of the energy of
the previous generation photon can be obtained by substituting
$\SUBB{\epsilon}{\pm}{\textsc{f}}$ into eq. (\ref{eq:e_RICS})
\begin{equation}
  \label{eq:eRICS-next-gen}
  \epsilonGRICS^{(i+1)}=\epsilonG^i\, b\left[ 1 + \left(\frac{\chiA}{b}\right)^2 \right]^{-1/2}\,.
\end{equation}
This equation describes the energy degradation in each cascade
generation for the RICS process.  In Fig.~\ref{fig:fEsynErics} we plot
the ratio of the characteristic energies of synchrotron and RICS
photons (given by eqs.~(\ref{eq:eRICS-next-gen}) and
(\ref{eq:eSYN-next-gen}) correspondingly) produced by pairs created by
the same parent photon as a function of the parent photon energy
$\epsilonG^{(i)}$ and magnetic field strength $B$.  The energy of RICS
photons are always smaller than the energy of synchrotron photons, and
for $B<10^{12}$~G significantly so.  Because of the much faster energy
degradation in the RICS process, cascade branches initiated by RICS
photons are in general shorter than those initiated by synchrotron
photons.  In later cascade generations RICS photons freely escape the
cascade zone while synchrotron photons emitted by the same particles
are absorbed, still splitting the energy of the parent photons into
pairs.  Therefore, RICS cascade branches should be in general less
efficient in splitting the energy than synchrotron ones.

\begin{figure}[t]
  \centering
  \includegraphics[clip,width=\columnwidth]{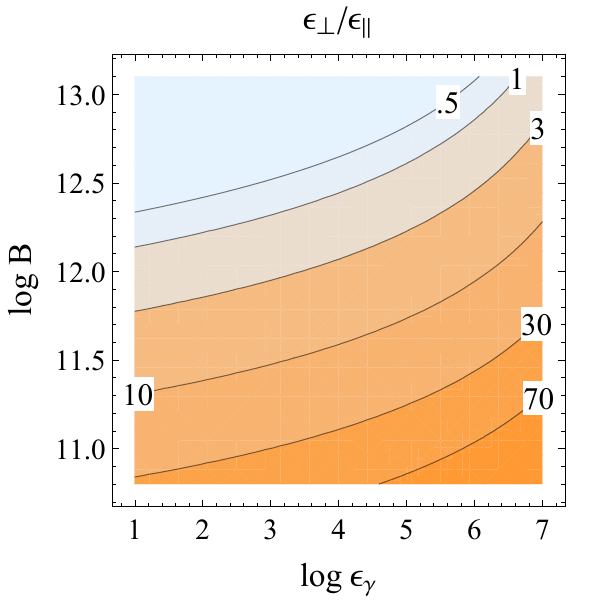}
  \caption{Ratio of perpendicular to parallel to $B$ energy of freshly
    created pairs: contours of $\epsilon_{\perp}/\epsilon_{\parallel}$
    as a function of logarithms of the parent photon energy
    $\epsilonG$ and magnetic field strength $B$ in Gauss for
    $\rhoC=10^7\mbox{cm}$. }
  \label{fig:fEperpEpar}
\end{figure}

Finally let us address the question of how much energy is going into
RICS branches of the cascade.  The energy powering RICS branches is
the kinetic energy of pairs moving along magnetic field lines, while
synchrotron branches are powered by the perpendicular energy of
freshly created pairs.  From
eqs.~(\ref{eq:gamma_F_chi}),~(\ref{eq:zeta_syn}) we get for the ratio
of the perpendicular to parallel pair energies
\begin{equation}
  \label{eq:e_perp2e_par}
  \frac{\epsilon_\perp}{\epsilon_\parallel}=
  \left[ 1 + \left(\frac{\chiA}{b}\right)^2 \right]^{-1/2}-1\,.
\end{equation}
As $\chiA>b$ (see eq.~(\ref{eq:chiA_limit_derivation})), the minimum
value for this ratio is $\sqrt2-1\simeq0.414$.  In
Fig.~\ref{fig:fEperpEpar} we plot $\epsilon_\perp/\epsilon_\parallel$
as a function of the energy of the parent photon and magnetic field
strength $B$.  For most of the parameter space in each pair creation
event, the fraction of energy going into a the RICS cascade branch is
smaller than those going into the synchrotron branch.  Even in the
case when more energy is left in the pairs' parallel motion, the
energy available to the RICS branch is only $1.5$ times larger than
the energy available for synchrotron branch.  In this case, however,
the value of $\chiA$ is close to $b$, i.e. pair formation occurs near
the kinematic threshold and is not very efficient (see
\S~\ref{sec:opacity-gamma-b}).  Hence, cascades where RICS branches
have more available energy should not be very efficient.

Summarizing the above arguments, RICS branches are less efficient in
splitting photon energy into pairs and the energy available to those
branches is in the best case comparable to the energy available for
synchrotron branches.  The effect of RICS branches are strongest for
higher values of the magnetic field,
$\gtrsim\mbox{few}\times10^{12}$~G, when pair formation happens close
to kinematic threshold and the cascade multiplicities are lower than
for weaker magnetic fields.  The final multiplicity of the total
cascades should be less than twice that of the pure CR-synchrotron
cascade in the best case and so the pure CR-synchrotron cascades
studied in this paper provide good estimates for the multiplicity of
strong polar cap cascades.

\section{Flux of primary particles and pair yield}
\label{sec:flux-primary}

\begin{figure}
  \includegraphics[clip,width=\columnwidth]{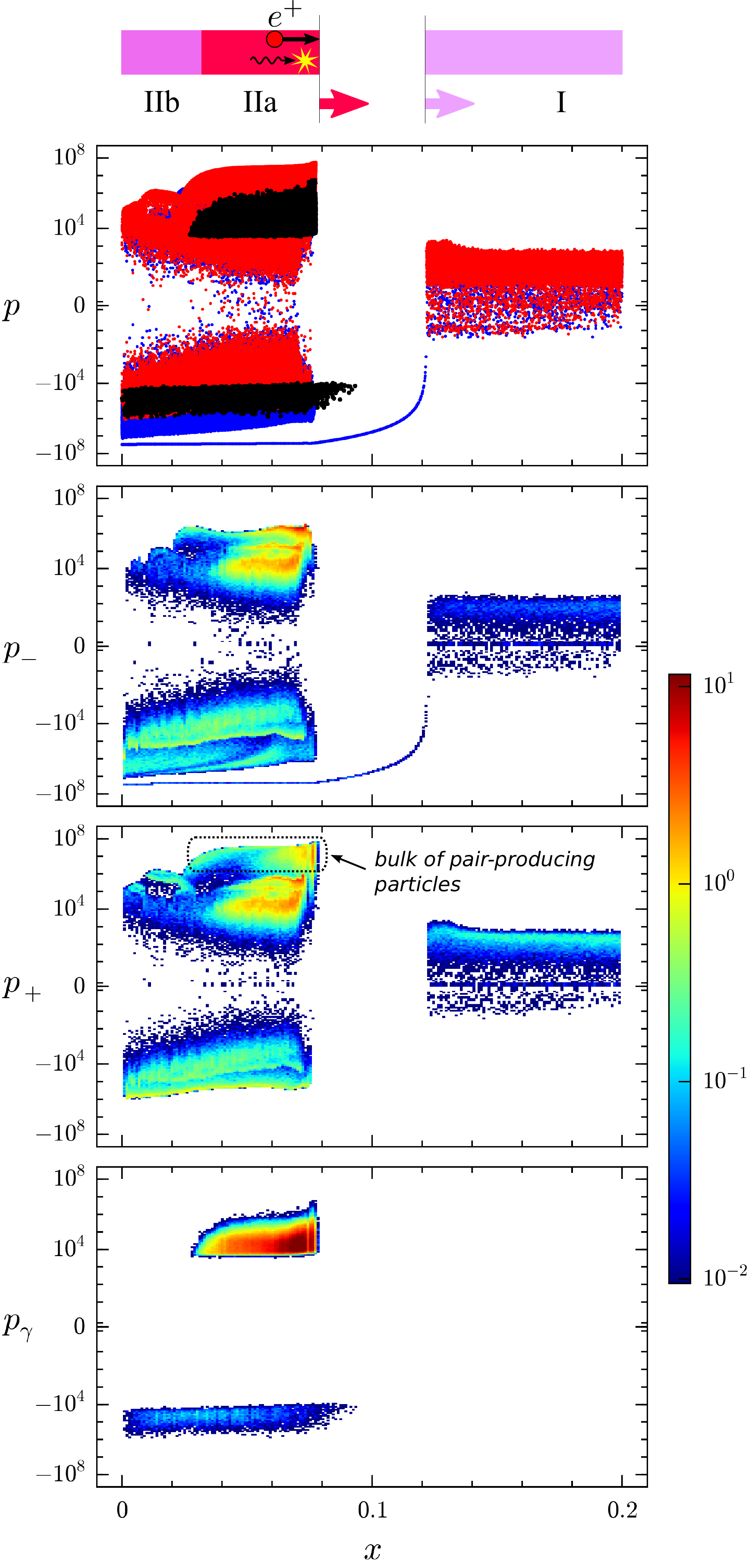}  
  \caption{Snapshot of the phase space for Ruderman-Sutherland cascade at the
    end of a discharge. Cascade an pulsar parameters: $\jm/\GJ{j}=1$
    $B=10^{12}$~G, $P=33$~ms, $\xi_j=1$. Horizontal axis -- particle positron
    $x$, vertical axis -- particle momentum normalized to $m_ec$; the vertical
    axis is logarithmic except for the region around zero momentum ($-5<p<5$),
    where the scale is linear.  The top panel shows phase space portrait of the
    cascade: each dot represents a numerical particle (every 10th particle is
    plotted); blue dots -- electrons, red dots -- positrons, black dots --
    photons.  Three plots beneath show particle number density in phase space:
    $p_{-}$~vs~$x$ -- electrons, $p_{+}$~vs~$x$ -- positrons,
    $p_{\gamma}$~vs~$x$ -- photons.  Particle number density is color-coded
    according to the color map on the right in units of $\GJ{n}$.  Particles
    which produce most of the pairs are positrons inside the area surrounded by
    dotted line marked as ``bulk of pair-producing particles''. On the top of
    the plots we show the sketch of the structure of the acceleration zone from
    Fig.~\ref{fig:accelerator}(b).}
  \label{fig:xp_xpdensity_RS}
\end{figure}
  
\begin{figure}
  \includegraphics[clip,width=\columnwidth]{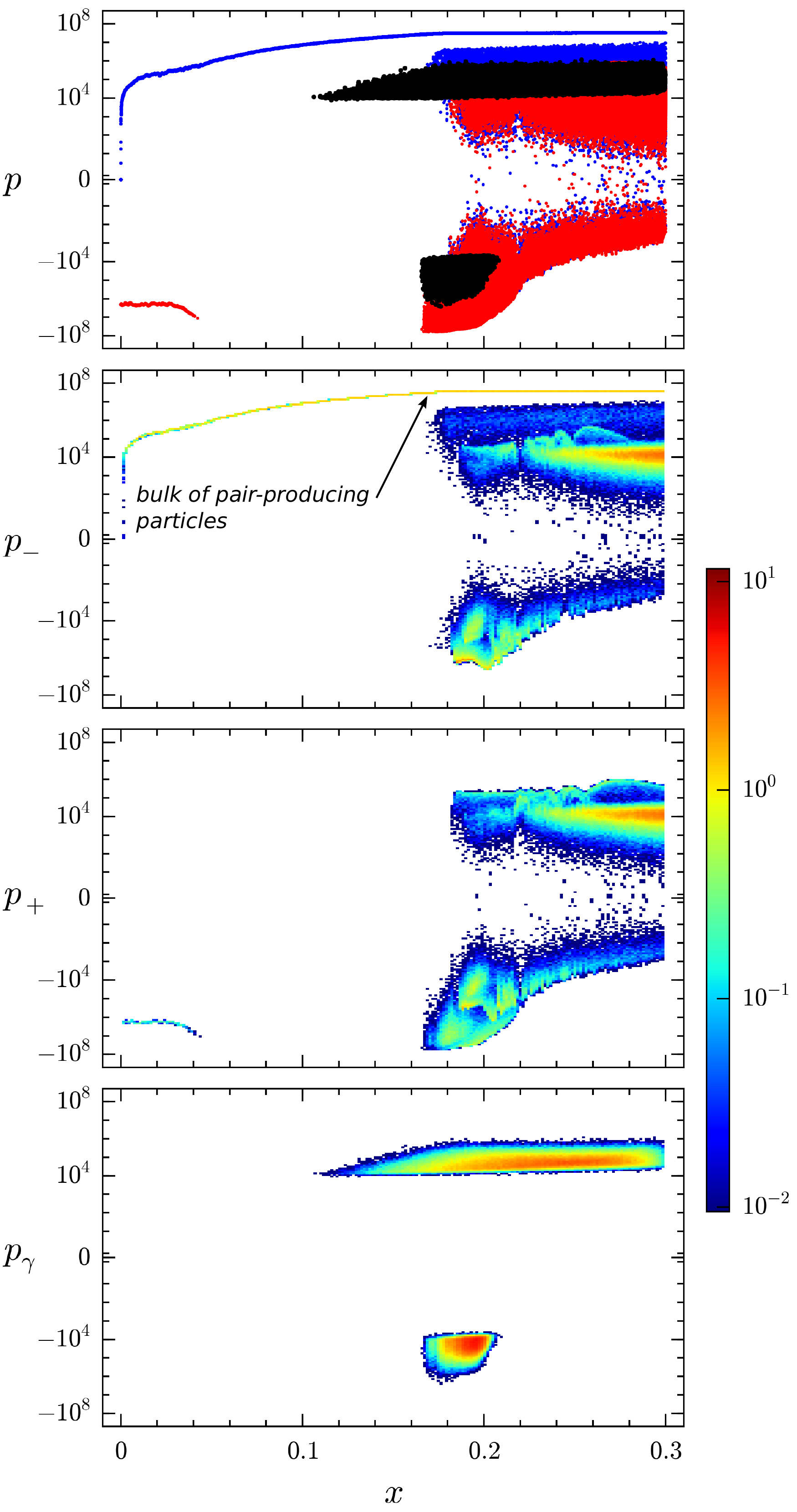}  
  \caption{Snapshot of the phase space for cascade in Space Charge
    Limited Flow regime. Cascade an pulsar parameters:
    $\jm/\GJ{j}=1.5$ $B=10^{12}$~G, $P=33$~ms, $\xi_j=0.25$. Types of
    plots and notations are the same as in
    Fig.~\ref{fig:xp_xpdensity_RS}.}
  \label{fig:xp_xpdensity_superGJ}
\end{figure}

As we have shown above, for the most efficient polar cap cascades the
total pair yield for a given value of the magnetic field $B$ depends
mostly on the flux of primary particles.  According to self-consistent
models of polar cap acceleration zones (T10, TA13), particle
acceleration is intermittent and the pattern of plasma flow and
acceleration efficiency depends on the ratio of the current density
imposed by the magnetosphere to the GJ current density $\jm/\GJ{j}$ as
well as the boundary conditions at the NS surface -- whether particles
can be extracted from the surface or not.

There are essentially three qualitatively different regimes of plasma
flow that determine the flux of primary pair-producing particles: (i)
in space charge limited flow with $0<\jm/\GJ{j}<1$ particles are
accelerated up to very low energies, $\epsilonP\sim$~few, and no pairs
are produced, (ii) in space charge limited flow with $\jm/\GJ{j}>1$ a
significant flux of primary particles of the order of $\sim\GJ{j}/e$
is accelerated through the gap while the gap is moving toward the NS,
(iii) in all other cases -- i.e. for any current density in
Ruderman-Sutherland model and for $\jm/\GJ{j}<0$ in space charge
limited flow regime -- in each burst of pair formation a blob of
primary particles is produced during formation of the gap, and then no
significant amount of primary particles is created until the formation
of the next gap.

In Fig.~\ref{fig:xp_xpdensity_RS} we plot as an example for case (iii)
a phase portrait and densities of plasma and photons in the phase
space for a Ruderman-Sutherland cascade with $\jm/\GJ{j}=1$.  The
process of gap formation for this case is described at the beginning
of \S\ref{sec:energy-primary}.  When the gap is formed, only a very
few particles leak from the plasma tail -- in this case electrons,
visible as a line-like feature in the top two panels of
Fig.~\ref{fig:xp_xpdensity_RS} -- are accelerated in the gap, the flux
of these particles is much less than $\sim\GJ{j}/e$ and their
contribution to the pair production can be neglected.  The blob of
primary particles which was created during the gap formation has a
density of a few $\GJ{n}$ in its densest parts and a size comparable
to the gap's height.  Those particles -- in this case positrons in the
area surrounded by a dotted line on the plot for positron phase space
density in Fig.~\ref{fig:xp_xpdensity_RS} -- create the vast majority
of pairs.  The density of primary particles in the blob is high but no
new primary particles are created until the next gap is formed, the
duty cycle of the cascade is small.  The resulting flux of primary
particles averaged over the period of gap formation is rather low.

In case (ii), in space-charge limited flow with super GJ current
density, the situation is qualitatively different in that during the
lifetime of the gap, and not only during the time the gap is forming,
as in case (iii), a significant constant flux of particles is going
through the acceleration zone.  The duty cycle of such a cascade
should be substantially higher than in any other types of cascades.
As an example of such a cascade we plot in
Fig. \ref{fig:xp_xpdensity_superGJ} a phase portrait and the densities
of plasma and photons in phase space for a cascade in space-charge
limited flow regime with $\jm/\GJ{j}=1.5$.  The gap is formed at some
distance from the NS and moves toward it.  When the first particles
are formed the process of electric field screening proceeds in a
similar way to the case of Ruderman-Sutherland cascades; the blob of
ultrarelativistic particles is created and particles leaking from it
create a tail of mildly relativistic plasma screening the electric
field behind the blob; this blog moves toward the NS creating pairs.
However a constant flux of particles extracted from the NS surface --
in this case electrons, visible as a line-like feature in the top two
panels of Fig.~\ref{fig:xp_xpdensity_superGJ} -- are accelerated in
the gap.  The density of these electrons is high (for the case shown
in Fig.~\ref{fig:xp_xpdensity_superGJ} it is $\simeq1.25\,\GJ{n}$) and
these particles, and not the particles from the blob, produce most of
the pairs.

Because of the intermittency of pair formation, the resulting
multiplicity must be adjusted by the relative fraction of time during
which primary particles are produced.  If $\SUB{\tau}{active}$ is the
time of active particle acceleration and $\SUB{T}{cascade}$ is the
time between the beginning of successive bursts of pair creation, then
the pair multiplicity as discussed in previous sections should be
multiplied by the attenuation factor
\begin{equation}
  \label{eq:f_kappa}
  f_\kappa=\frac{\SUB{\tau}{active}}{\SUB{T}{cascade}}
\end{equation}
to get the average multiplicity of pair cascades.

The existing self-consistent simulations of particle acceleration
(T10, TA13) are inconclusive about the cascade repetition rate.  The
numerical resolution in these simulations was inadequate for that
purpose -- because of copious pair formation the Debye length of
plasma at some point became smaller than the cell size and the
formation of the plasma tail could not be simulated accurately; the
repetition rate, however, is determined by the mildly relativistic
plasma in the plasma tail, as the next burst of pair formation starts
only when plasma leaves the polar cap region.  Future numerical
simulations would address this issue and provide accurate values for
cascade duty cycles; meanwhile here we will try to give a rough
estimate for the cascade repetition rate using the following simple
physical arguments.

The strongest pair formation can continue up to the distance of the
order of a few NS radii $\RNS$.  Although the exact physical mechanism
responsible for formation of the mildly relativistic plasma tail is
not clear (see previous paragraph) it seems to us plausible to expect
that when plasma injection due to pair formation stops, reversal of
some pairs toward the NS -- plasma leakage from the blob to the tail
-- should taper off as well.  So, the time when plasma is filling the
polar cap region is of the order of $\sim\RNS/c$.  The mildly
relativistic pairs which fill the polar cap region after the blob of
plasma has moved away are moving with relativistic velocities, so the
time the plasma needs to clear out is also of the order of
$\sim\RNS/c$.  Therefore, the cascade repetition rate seems to be
$T_{cascade}\sim\mbox{few}\times\RNS/c$

For the case (iii), in the flow regime with vacuum gaps --
Ruderman-Sutherland cascades or space-charge limited flow in return
current regions -- particles are accelerated only during formation of
the gap, the time of active particle acceleration is quite short.  The
longitudinal size of the blob of primary particles is about the size
of the gap $\GAP{h}$, and the primary particle density in the blob is
a few $\GJ{n}$.  All primary particles in the blob move in the same
direction and the blob passes any given surface normal to the field
lines during time $\SUB{\tau}{active}\sim{}\GAP{h}/c$. If the next gap
will form after the blob of primary particles moves a few $\RNS$ away,
the attenuation factor should be
\begin{equation}
  \label{eq:f_kappa_RS}
  f_\kappa\sim\frac{\GAP{h}}{\RNS}\simeq
  \sci{4}{-2}\: \chiA^{1/7} \xi_j^{-3/7} \SUB{\rho}{c,\,7}^{2/7} P^{3/7} B_{12}^{-4/7}
\end{equation}
For the Crab pulsar this attenuation factor is around $0.001$ if we
assume $P=33$~ms, $B=\sci{3}{12}$~G, $\chiA=1/7$, and $\xi_j=1$ (for
inclination angle $60^\circ$ and current density $\jm\simeq\GJ{j}$).
The resulting multiplicity of such cascades in Crab would be
$\lesssim{}10^3\GJ{n}$.

In the space-charge limited flow regime with super-GJ current density,
case (ii), the cascade efficiency should be much higher.  The gap
appears at some distance and moves toward the NS (see \S~6.2 in TA13).
The process of gap formation is very similar to that in cascades for
case (iii) discussed above.  A blob of relativistic particles is
created, which in this case propagates toward the NS, and pair plasma
leaking from it creates a plasma tail filling the region behind the
blob which prevents formation of the next gap.  But in this case
particles extracted from the NS surface flow into the magnetosphere
through the gap and are accelerated all the time the gap exists; these
particles create most of the pairs.  When the gap disappears, particle
acceleration stops.  The remaining plasma should leave the gap and
flow into the magnetosphere before the next gap can appear. The time
necessary for the remaining plasma to leave should be comparable to
the time of the gap's existence, as all particles are
relativistic. Therefore, in this case the attenuation factor should be
$f_\kappa\sim{}1/\mbox{few}$ -- much higher than given by
eq.~(\ref{eq:f_kappa_RS}). So, polar cap regions with super-GJ current
density should have a pair yield of about $10^5\GJ{n}$.

Let us summarize our findings.  Pulsars with polar cap cascades
operating in the Ruderman-Sutherland regime would not be efficient
pair producer, the highest yield should be less that
$\sim{}10^3\GJ{n}$.  If the polar cap cascades operate in the space
charge limited flow regime, then in regions with super-GJ current
densities the pair yield is quite high, around $10^5\GJ{n}$; in the
regions with return current (anti-GJ current density) the yield should
be less that $\sim{}10^3\GJ{n}$, and in regions with sub-GJ current
densities no pair plasma would be produced at all.  We therefore
conclude that the maximum pair yield of a young pulsar would be less
that $10^5\GJ{n}$, as only a fraction of the polar cap can have
super-GJ current density.

Pair yield is mostly affected by the duty cycle of the cascade, and
not by pulsar parameters such as magnetic field strength $B$, radius
of curvature of magnetic field lines $\rhoC$, or pulsar period $P$.
This holds only for the regime in which the gap height is much smaller
than the PC radius, true for all young pulsars.  When the gap height
approaches and exceeds the PC radius, the pair yield drops quickly, as
shown in Figure~\ref{fig:multiplicity}.  The distribution of
$\jm/\GJ{j}$ in the polar cap is determined by the pulsar inclination
angle and we expect that inclination angle should be the most
important factor determining pair yield of a young pulsar.

\section{Discussion}
\label{sec:discussion}

We have performed a systematic study of electron-positron pair
cascades above pulsar polar caps for a variety of input parameters
including surface magnetic field, pulsar rotation period, primary
particle energy and magnetic field radius of curvature.  We have also
studied here for the first time the multiplicity of self-consistent
pair cascades, i.e. those that are capable of generating currents
consistent with global magnetosphere models.

We find that pair multiplicity is maximized for a magnetic field
strength near $10^{12}$~G, independent of the other parameters.  This
value of field strength strikes a balance between maximizing the
fraction of photons that pair produce -- the weaker the magnetic field
the higher the number of escaping photons -- and the fraction of
photon energy going into pair production -- in stronger fields photons
are absorbed when they have smaller angles to $B$, and created pairs,
having smaller perpendicular to $B$ momenta, emit less synchrotron
photons, which produce the next generation of pairs.  This is true for
curvature radiation - synchrotron cascades, which should be the
dominant source of pairs in young energetic pulsars. For stronger
magnetic fields $B\gtrsim\sci{3}{12}$~G resonant inverse Compton
scattering of soft X-ray photons emitted by the NS surface can tap
some of the pair's energy parallel to $B$ and increase the cascade
multiplicity.  RICS cascade branches, however, are less efficient than
the synchrotron branches; this together with the decrease of the
photon absorption cross-section for near threshold pair creation in
strong magnetic fields should not change the fact the polar cap
cascade multiplicity reaches its maximum for $B\sim10^{12}$~G.

We find that the pair multiplicity at the peak of the cascade cycle,
$\kappa\sim10^5$, is remarkably insensitive to pulsar period, magnetic
field and radius of curvature of magnetic field lines.  The reason for
this is self-regulation of the accelerator by pair creation: for
pulsar parameters resulting in more efficient pair production, the
size of the acceleration gap is smaller and the primary particle
energy is lower and vice-versa.  The most important factor determining
the multiplicity of polar cap cascades is the flux of primary
particles which depends on the ``duty cycle'' of the particle
acceleration in time-dependent cascades.  Estimating the ``duty
cycle'' we find that the time-averaged pair multiplicity is limited to
$\sim 10^{5}$ for the case of space charge-limited flow (free particle
extraction from the NS) with super Goldreich-Julian current density
($j/\GJ{j}>1$) and to only $\sim 10^{3}$ for the case of
Ruderman-Sutherland gaps (no particle extraction from the PC) and
space charge-limited flow with anti Goldreich-Julian current density
($j/\GJ{j}<0$).  The current density distribution in the PC depends on
the pulsar inclination angle, and so for young energetic pulsars the
inclination angle should be the most important factor determining the
total multiplicity of plasma in the magnetosphere.

Our finding of a peak multiplicity from self-consistent pair cascades
in young pulsars ($\sim 10^5$), that is largely independent of their
period and magnetic fields, is very different from previous
predictions from steady cascades that produce a range of
multiplicities ($10^2 - 10^4$) that depend strongly on pulsar
parameters.  This difference primarily results from much smaller gap
heights in the time-dependent cascades, that keep the gap size much
less than the PC radius over a large parameter range.  This in turn
results in a nearly uniform magnetic field strength throughout the
gaps, higher accelerating electric fields, higher primary particle
energies and more efficient pair production -- primary particles are
accelerated faster and photons are injected in the region of stronger
magnetic field than in previous steady cascade models.

We have also numerically simulated the photon spectra from
self-consistent polar cap pair cascades.  Our results show that the
cascade photon spectra from the more compact gaps nearer to the
neutron star surface have cutoffs that are around 10 - 100 MeV.  This
is at the lower end the \textsl{Fermi} energy band ($30$~MeV -
$300$~GeV), which may explain why \textsl{Fermi} has not seen strong
evidence for $\gamma$-ray emission from the PC.  High energy emission
from pair cascades is thus expected to occur in lower energy bands,
below 100 MeV.  PC cascade emission may have been detected recently
from PSR J1813-1246 \citep{Marelli2014} at X-ray energies by
XMM-Newton and Chandra.  The unusual X-ray light curve that shows two
peaks separated by 0.5 in phase but offset by a quarter of a period
with the $\gamma$-ray peaks can be explained through a geometric model
placing the $\gamma$-ray emission in the outer magnetosphere and the
X-ray emission at lower altitude above the PCs, and could be coming
from the PC cascades.

Our results show that PC cascades are more efficient in producing
electron-positron pairs that was previously assumed.  However, even
the higher pair multiplicity of $10^5$ is not enough to account for
the fluxes of particles needed to explain the synchrotron radiation
from PWNe.  Based on a multiplicity of $10^5$, we estimate that the
Crab pulsar produces a pair flux from each PC of about
$\sci{2}{39}\,\mbox{pairs}\,\mbox{s}^{-1}$.  The flux from both PCs is
an order of magnitude smaller than the pair flux required to account
for the radiation from the nebula, which is estimated to be about
$\sim \sci{4}{40}\,\mbox{pairs}\,\mbox{s}^{-1}$ \citep{deJager1996}.
Cascades in the outer magnetosphere are not very efficient pair
producers \citep[e.g.][]{Hirotani2006}, and so the injection of plasma
by pulsars can not account for the population of particles in PWNe
emitting at radio wavelengths.  These radio emitting particles must
then have a different origin from particles emitting at shorter
wavelength; for example, they might be picked up from the gas
filaments in the supernova remnant or be remnants of some unknown
acceleration mechanism in the early history of the nebula
\citep{AtoyanAharonian1996}.

Another implication of our results concerns observations of cosmic-ray
electrons and positrons at Earth that have shown an excess of
positrons over what can be produced in secondary cosmic-ray
interactions \citep{Adriani2009,Accardo2014}, indicating the existence
of primary positron sources in the Galaxy.  Various studies
\citep[e.g.][]{Gendelev2010} have estimated that PWNe could account
for the excess of cosmic ray positrons.  With higher pair multiplicity
from young pulsars, and subsequent acceleration of the pairs at the
pulsar wind termination shock, PWNe could produce a more significant
contribution of primary cosmic ray positrons.

\acknowledgments

This work was supported by a NASA Astrophysics Theory grant and a
\textsl{Fermi} Cycle-5 guest investigator grant.

\appendix

\section{Algorithms for semi-analytical calculation of cascade
  multiplicity}
\label{sec:algorithms}

Here we show pseudo-codes of algorithms used to compute cascade
multiplicity.  For calculation of $\chiA(\epsilonG,B,\rhoC)$ we
computed and stored a table of $1/\chiA$ values for a uniformly
divided grid $77\times30\times20$ in
$\log\epsilonG\times\log{B}\times\log{\rhoC}$ space, and then used
cubic piece-polynomial interpolation to get $\chiA$ for parameter
values required by expressions used in the algorithms.

Algorithm~\ref{alg:sync_mult} computes total number of particles
produced in synchrotron cascade initiated by a single primary photon
with the energy $\epsilonG$, physical processes are described in
\S\ref{sec:mult-synchr-casc}. \texttt{e\_phot} and \texttt{n\_phot}
are the number and the energy of synchrotron photons emitted by
particles of the current generation, \texttt{e\_phot} is the energy of
escaping photons, and \texttt{n} is the total number of particles
produced in all previous generations.

\begin{algorithm}
  \label{alg:sync_mult}
  \SetKwProg{Fn}{Function}{:}{end}
  \SetKwFunction{Nsynch}{$\SYN{N}$}
  \SetKwData{vN}{n}
  \SetKwData{vNphot}{n\_phot}
  \SetKwData{vEps}{e\_phot}
  \SetKwData{vEpsEsc}{e\_esc}
  \KwData{$\epsilonG$ -- energy of primary photon,
    $\ESC{s}$ -- mfp of escaping photon }
  \KwResult{$\SYN{N}$ -- total number of particles produced in
    synchrotron cascade initiated by a photon with the energy $\epsilonG$}
  \BlankLine
  \Fn{\Nsynch{$\epsilonG,\ESC{s}$}}{
    $\vEpsEsc \leftarrow \epsilonGesc(\ESC{s})$\tcp*{eq.~(\ref{eq:eps_esc_eq})}
    $\vEps\leftarrow \epsilonG$\;
    $\vNphot \leftarrow 1$\;
    $\vN \leftarrow 0$\;
    \While{\vEps$\ge\vEpsEsc$}{
      $\vN \leftarrow \vN + 2~\vNphot$\;
      $\vNphot \leftarrow \vNphot~\SYN{n}(\vEps)$ \tcp*{eq.~(\ref{eq:n_sync})}
      $\vEps \leftarrow \epsilonG^{(i+1)}(\vEps)$ \tcp*{eq.~(\ref{eq:eSYN-next-gen})}
    }
    \KwRet{\vN}
  }
  \caption{Multiplicity of synchrotron cascade}
\end{algorithm}

Algorithm~\ref{alg:total_mult} computes the total multiplicity
$\kappa$ of a CR-synchrotron cascade according to
eq.~(\ref{eq:kappa-CR-Sync}). \texttt{e\_part} is the energy of the
primary particle at the current distance $s_i$; \texttt{e\_phot} is
the energy and \texttt{n\_phot} is the number of CR photons emitted by
the primary particle at the current distance, \texttt{Nsyn} is the
number of particles generated in synchrotron cascades initiated by the
primary particle at the current distance.  Integration over the
distance is done with the simple trapezoidal rule.

\begin{algorithm}[h]
  \label{alg:total_mult}
  \SetKwProg{Fn}{Function}{:}{end}
  \SetKwFunction{Nsynch}{$\SYN{N}$}
  \SetKwData{vN}{n}
  \SetKwData{vDN}{Nsyn}
  \SetKwData{vDNlast}{Nsyn\_last}
  \SetKwData{vS}{$s_i$}
  \SetKwData{vSlast}{$s_{i-1}$}
  \SetKwData{vNcr}{n\_phot}
  \SetKwData{vEphot}{e\_phot}
  \SetKwData{vEpart}{e\_part}
  \SetKwData{vImax}{i\_max}
  \KwData{$\epsilonP^0$ -- energy of the primary particle,
    $\CR{s}$ -- characteristic size of the cascade, 
    $\ESC{s}$ -- mfp of escaping photons }
  \KwResult{total number of pairs produced by the primary particle
    with the energy $\epsilonP^0$}
  \BlankLine  
  \Begin{
    divide $[0,\CR{s}]$ in subintervals $s_0,\dots{},s_{i_{max}}$\;
    $\vDNlast\leftarrow{0}$\;
    $\vN\leftarrow{0}$\;

    \For{$i\leftarrow 1$ \KwTo $i_{max}$}{
      \tcp{CR radiation at distance \vS}
      $\vEpart\leftarrow\epsilon_\pm(\epsilon^0_\pm, \vS)$ \tcp*{eq.~(\ref{eq:CR_es})}
      $\vEphot\leftarrow\epsilonGCR(\vEpart)$          \tcp*{eq.~(\ref{eq:CR_ephot})}
      $\vNcr\leftarrow\CR{n}(\vEpart)$                   \tcp*{eq.~(\ref{eq:n_CR})}

      \BlankLine
      \tcp{synchrotron cascade multiplicity}
      $\vDN \leftarrow \vNcr$ \Nsynch{\vEphot, $\ESC{s}$}\tcp*{algorithm~\ref{alg:sync_mult}}

      \BlankLine
      \tcp{trapezoidal rule}
      $\vN \leftarrow \vN + 0.5 (\vDN + \vDNlast) (\vS - \vSlast)$\;
      \vDNlast $\leftarrow$ \vDN\;
    }
    \KwRet{\vN}
  }
  \caption{Multiplicity of CR-synchrotron cascade}
\end{algorithm}

\section{Transition to radiation reaction limited regime}
\label{sec:trans-radi-react}

Radiation reaction begins playing an important role in the dynamics of
particle acceleration if particle energy losses per unit time become
comparable to the work done on the particle by the accelerating
electric field. For curvature radiation the condition for
applicability of the free acceleration regime is (see
eq.~(\ref{eq:CR_energy_losses}))
\begin{equation}
  \label{eq:rr_definition}
  eEc \gtrsim \frac{2}{3}\frac{e^2}{c^3} \left(
    \frac{c^2}{\rhoC}\right)^2 \epsilonP^4\,.
\end{equation}
The electric field $E$ grows linearly with the distance $\lP$ traveled
by the particle in the gap, eq.~(\ref{eq:E_xp}); the particle energy
$\epsilonP$ increases as $\lP^2$, eq.~(\ref{eq:epsilonP_s}), and at
some distance the condition~(\ref{eq:rr_definition}) is violated.
However, if condition~(\ref{eq:rr_definition}) holds up to the
distance $\lPe$ where the particle emits ``gap-terminating'' photons,
then the gap length and the final particle energies can be well
described by the free acceleration regime, neglecting radiation energy
losses.  Using eq.~(\ref{eq:E_xp}) for $E$, eq.~(\ref{eq:epsilonP_s})
for $\epsilonP$, and substituting $\lPeGAP$ for $\lP$ from the
eq.~(\ref{eq:s_e_free}) inequality~(\ref{eq:rr_definition}) becomes
the condition on the magnetic field strength
\begin{equation}
  \label{eq:B_rr}
  B > \frac{8}{3} \alphaF \chiA B_q \simeq \sci{2.8}{-3} B_q = \sci{1.15}{11}\mbox{G}\,.
\end{equation}
In the second step we used the value $\chiA=1/7$.  The radius of
curvature $\rhoC$ and pulsar period $P$ cancel out, except for a weak
dependence of $\chiA$ on $\rhoC$.  We see that for most
non-millisecond pulsars radiation reaction can be neglected.

\section{Limit on 1-d approximation}
\label{sec:limit-1-d}

The one-dimensional approximation for particle acceleration in the
gaps works well if the length of the gap is much smaller that the
width of the polar cap. As the length of the gap is $\sim{}2$ times
larger than $\GAP{l}$, we can set the formal limit on 1-D
approximation as
\begin{equation}
  \label{eq:lgap_smaller_rpc}
  \GAP{l}  < \PC{r}\,,
\end{equation}
where $\PC{r}=\sqrt{2\pi\RNS^3/Pc}$ is the polar cap radius. Using
eq.~(\ref{eq:l_gap}),~(\ref{eq:s_e_free}) for $\GAP{l}$ and $\lPe$ we
get the following limits on pulsar parameters for applicability of the
1-D approximation
\begin{equation}
  \label{eq:1d-limit-aux}
  B^{4/7}P^{-13/14}>\frac{7}{6}\left(\frac{c}{2\pi\RNS^3}\right)^{1/2}
    \left(\chiA\frac{B_q^4\lambdaC^2c^3}{\pi^3}\right)^{1/7}
    \xi_j^{-3/7}\, \rhoC^{2/7}\,.
\end{equation}
The pulsar magnetic field is usually estimated assuming magnetodipolar
energy losses from the values of period $P$ and period derivative
$\dot{P}$ as $B=\sci{3.2}{19}\sqrt{P\dot{P}}$. Expressing $B$ in this
way and putting the numerical values for physical constants, we get
from eq.~(\ref{eq:1d-limit-aux}) the following condition for
applicability of 1-D approximation in terms of $P$ and $\dot{P}$
\begin{equation}
  \label{eq:1d-limit}
  \dot{P} > \sci{3.21}{-22} \chiA^{1/2}\xi_j^{-3/2}\rhoC\, P^{9/4}
  \simeq \sci{4.3}{-16} \SUB{\rho}{c,\,7}\, P^{9/4}\,,
\end{equation}
in the last step we used values $\xi_j=2$ and $\chiA=1/7$. For
cascades along dipole magnetic field lines at the edge of the polar
cap with the radius of curvature
\begin{equation}
\rhoC=\frac{4}{3}\frac{\RNS}{\PC{\theta}}\approx{}\sci{9.2}{7}\sqrt{P}\,\mbox{cm}, 
\label{eq:rhoC-dipole-PCedge}
\end{equation}
where $\PC{\theta}=\sqrt{2\pi\RNS/cP}$ is the colatitude of the polar
cap edge, condition~(\ref{eq:1d-limit}) takes the form
\begin{equation}
  \label{eq:1d-limit-dipole-PC_edge}
  \dot{P} >  \sci{4}{-16}  P^{11/4}\,.
\end{equation}

\bibliographystyle{apj} 
\bibliography{fmc_1,pulsars_theory,pulsars_obs}

\end{document}